\documentstyle[12pt,epsfig]{article} 
\newlength{\dinwidth} 
\newlength{\dinmargin} 
\def\parallel{| \hskip-0.03cm |}
 
\newcommand{\ba}{\begin{array}} 
\newcommand{\ea}{\end{array}} 
\newcommand{\be}{\begin{equation}} 
\newcommand{\ee}{\end{equation}} 
\newcommand{\bee}{\begin{eqnarray}} 
\newcommand{\eee}{\end{eqnarray}} 

\begin{document} 
\thispagestyle{empty} 
\addtocounter{page}{-1} 
\begin{flushright} 
SNUST 99-001\\
UOSTP 99-003\\ 
{\tt hep-th/9902101}\\ 
\end{flushright} 
\vspace*{1.3cm} 
\centerline{\Large \bf Holographic View of Causality and Locality}
\vskip0.4cm
\centerline{\Large \bf via}
\vskip0.4cm
\centerline{\Large \bf Branes in AdS/CFT Correspondence~\footnote{ 
Work supported in part by KOSEF Interdisciplinary Research Grant, 
KRF International Collaboration Grant, Ministry of Education Grant 
1998-015-D00054, UOS Academic Research Program, SNU 
Acacemic Research Grant, and The Korea Foundation for Advanced Studies 
Faculty Fellowship. }} 
\vspace*{1.2cm} 
\centerline{\large \bf Dongsu Bak${}^1$ $\quad$ \& $\quad$ Soo-Jong Rey${}^2$
} 
\vspace*{0.8cm} 
\centerline{\it Physics Department,  
University of Seoul, Seoul 130-743 Korea${}^1$} 
\vskip0.3cm 
\centerline{\it Physics Department, Seoul National University, 
Seoul 151-742 Korea${}^2$} 
\vskip0.6cm 
\centerline{\tt dsbak@mach.uos.ac.kr, \quad sjrey@gravity.snu.ac.kr } 
\vspace*{2.5cm} 
\centerline{\Large\bf abstract} 
\vspace*{0.5cm} 
We study dynamical aspects of holographic correspondence between $d=5$ 
anti-de Sitter supergravity and $d=4$ super Yang-Mills theory. We probe 
causality and locality of ambient spacetime from super Yang-Mills theory by 
studying transmission of low-energy brane waves via an open string stretched 
between two D3-branes in Coulomb branch. By analyzing two relevant physical 
threshold scales, we find that causality and locality is encoded in the super 
Yang-Mills theory provided infinite tower of long supermultiplet operators are 
added. Massive W-boson and dual magnetic monopole behave more properly as 
extended, bilocal objects. We also study causal time-delay of low-energy 
excitation on heavy quark or meson and find an excellent agreement between 
anti-de Sitter supergravity and super Yang-Mills theory descriptions. We 
observe that strong `t Hooft coupling dynamics and holographic scale-size 
relation thereof play a crucial role to the agreement of dynamical processes.

\vspace*{1.1cm} 
 
\baselineskip=19pt 
\newpage 
 
\section{Introduction} 
\setcounter{equation}{0}
It is now widely believed that an anti-de Sitter (AdS) supergravity is dual 
to large N limit of conformal field theory (CFT) \cite{maldacenaproposal}. 
A particularly interesting case of this so-called AdS/CFT correspondence 
is between semiclassical limit of Type IIB superstring compactified to 
five-dimensional anti-de Sitter spacetime and large-$N$ limit of super 
Yang-Mills theories in four dimensions. As exemplified by this case, 
one of the most important features is that
the `holographic principle' \cite{thooft, susskind, thorn} underlies
all AdS/CFT correspondences: semi-classical gravity in $(d+1)$ dimensions
is holographically is described by degrees of freedom of 
large-$N$ quantum field theory in $d$-dimensions.
The AdS/CFT correspondence has been test extensively and evidence for 
it has been drawn mainly from agreement between the two sides of 
various {\sl static} quantities, such as
symmetries, particle spectrum, correlation functions, 
and black hole entropy \cite{gubserklebanovpolyakov, witten, susskindwitten}. 

The holographic principle, however, refers to the correspondence not just 
for static properties only but also for all {\sl dynamical} ones \footnote{
Minkowski description of AdS/CFT correspondence has been studied
extensively in \cite{bala}. Various dynamical issues on anti-de Sitter 
spacetime has been address in \cite{banksetal, polchinskipeet, horowitzitzhaki, polchinskiflat, susskindflat, kabatlifschytz}.}. This raises
an immediate puzzle: how can possibly local and causal dynamics on anti-de 
Sitter spacetime be recovered from a lower-dimensional quantum field theory? 
Large-$N$ super Yang-Mills theory is certainly a well-defined quantum field 
theory \cite{thooftlargen} with locality and causality of four-dimensional
Minkowski spacetime built in. So is the supergravity in anti-de Sitter 
spacetime, in the semi-classical limit at the 
least. Given the apparent locality and causality of physics, it seems
virtually impossible to describe what happens in the bulk of spacetime by
dynamical variables located at the boundary. 
In this paper, taking the correspondence between Type IIB superstring
on $d=5$ anti-de Sitter spacetime and $d=4$ super Yang-Mills theory as a 
setup, we will try to provide an answer to this puzzle.  

More specifically, we will pose and study the following question: in order to 
encode {\sl causality} and {\sl locality} of semi-classical supergravity in 
$d=5$ anti-de sitter spacetime, what precise prescription does one need in 
formulating the $d=4$ super Yang-Mills theory? Our main conclusion will be 
that neither conventional (keeping dimension-four operators only) nor 
Dirac-Born-Infeld form
is sufficient and only after infinite tower of higher-dimensional operators 
encompassing {\sl long supermultiplets} are included, causality and locality
can be decoded out of the super Yang-Mills theory.   

We will address the posed question in the Coulomb branch of the super
Yang-Mills theory, viz. consider $SU(N) \rightarrow S[U(N-2) \times
(U(1) \times U(2))]$ by separating two D3-branes out of the $(N-2)$ remainder
and place them at $r_{1,2}$, respectively, in the Coulomb branch moduli space.
We will then place a macroscopic open string (either fundamental or Dirichlet) 
stretched between the two D3-branes and study transmission of localized
brane waves from the first D3-brane to the second through the string. The
stretched open string is identified with a massive W-boson with mass
$M_{\rm w} = (r_1 - r_2) \equiv \Delta r$. As the string is stretched over a 
distance $\Delta r$, under suitable circumstance, one would expect to 
detect causal time-delay out of the transmitted brane-waves. 

This process, which is a sort of Thomson scattering \cite{jackson} 
(alias Compton scattering at low-energy) of brane waves off a 
massive W-boson, is closely related to much-studied, $F^4$ brane-wave 
interaction \cite{f4, dineseiberg, maldacenaf4, periwal, berkooz, lowe}. 
Cutting the one-loop diagram of the latter across
two internal W-boson lines, one obtains a pair of tree-level, Thomson 
scattering diagram. In so far as probing causality and locality, the only
difference between the two would be that the $F^4$ brane-wave 
interaction describes inter-brane interaction via exchange of 
closed string states, while the Thomson scattering interaction describes
interaction via virtual excitation of the massive W-boson. In effect, the
stretched open string serves as a sort of one-dimensional `waveguide' of 
the radiation, which would otherwise have propagated out all over ambient 
spacetime.

Comparison of Thomson scattering with known results regarding $F^4$ brane-wave
interaction brings up another important lesson drawn throughout this work: 
{\sl static} 
and {\sl dynamic} information are closely related in flat spacetime,
but {\sl not} in anti-de Sitter spacetime. 
In flat spacetime (corresponding to $N=2$), it is 
well-known that the $F^4$ brane-wave interaction is exact at one-loop by
the non-renormalization theorem of \cite{dineseiberg}. Douglas and 
Taylor \cite{douglastaylor}
have argued that, by taking a coincident $(N-2)$-brane limit from a generic
point in the Coulomb branch, the non-renormalization theorem should also hold 
in anti-de Sitter spacetime. As an evidence for this assertion, they have
noted that, the supergravity interaction for zero-momentum brane waves is the
same for both flat and anti-de Sitter spacetime. Through careful analysis, 
Das \cite{das} has confirmed the result further \footnote{ 
Intuitively, the result may
be interpreted as a consequence of the fact that operators involved are short 
supermultiplets and that the massive W-boson running through the loop is a 
BPS state.}. The interaction at zero-momentum, however, is always 
instantaneous and retarded interaction will show up only if nonzero 
momentum configuration is considered, a well-known fact of nonrelativistic
expansion in covariant perturbation theory. 
For brane waves with nonzero momentum, corrections to the leading
order result in Yang-Mills theory is controlled by the string scale, $\ell_s$
\cite{douglastaylor},
hinting that stringy corrections somehow ought to be taken into account to 
the supergravity description. Douglas and Taylor have interpreted 
these corrections as a manifestation of 
the spacetime uncertainty principle \cite{yoneya}. Das
has sketched calculation of retarded interaction between nonzero momentum
brane waves in flat spacetime and have argued that the non-renormalization 
theorem assures the same result in anti-de Sitter spacetime. Underlying to 
their arguments seems that the massive W-boson, which is a stretched open 
string of length $\Delta r$, is a good ruler for measuring distance and hence
causality of the bulk spacetime, both flat and anti-de Sitter. 

From Thomson scattering process, we will find contrasting results. Underlying
to the process, the W-boson, being an open string stretched between the
two D3-branes, sets two competing threshold scales: winding and oscillation
energy of the string. We will find that causality and locality of the ambient
spacetime is visible in super Yang-Mills theory only if the oscillation 
threshold is lighter than the winding threshold. The winding threshold remains
the same for both flat and anti-de Sitter spacetime but, quite importantly, 
the oscillation threshold turns out to change significantly \footnote{
Recall that a winding string is a BPS state, but not when oscillator 
excitations are added. }. 
In flat spacetime, crossover between the two threshold takes place
when $\Delta r \sim {\cal O}(\ell_s)$, confirming from dynamical side 
the well-known result on D-brane worldvolume dynamics  \cite{polchinski}. 
In anti-de Sitter spacetime, however,  crossover takes place
when (using Maldacena's scaling variable $U \equiv r / \ell_s^2$)
\be
{(\Delta U)^2 \over U_1 U_2} \qquad \sim \qquad 
{\cal O}\left({1 \over g_{\rm st} N} \right). 
\ee
It shows that, everywhere in the Coulomb branch (possibly except at the 
origin), stong `t Hooft coupling limit ensures that the oscillation threshold 
is lower than the winding threshold and hence enables the large-$N$ super 
Yang-Mills theory to encode causality and locality of semi-classical 
supergravity in the anti-de Sitter spacetime.  

With light oscillation threshold, the super Yang-Mills theory is given in a
rather different form from conventional (retaining dimension-four operators 
only) or Dirac-Born-Infeld theories. 
The massive W-bosons and dual magnetic monopoles are more suitably described
as a sort of extended, bilocal object \footnote{rather reminiscent of Yukawa's
old idea \cite{yukawa}.}. Remarkably, particle Lagrangian of the W-boson takes
(an extended form of) Pais-Uhlenbeck \cite{paisuhlenbeck} model with 
transcendental, infinite-order kernels, which is one of the few known 
examples satisfying convergence, positive-definiteness and strict causality. 

This paper is organized as follows. In section 2, we will study brane-wave
transmission via Thomson scattering, first in flat spacetime. In section 3,
we will study the holographic relation between energy and size in AdS/CFT 
correspondence, but now from dynamical point of view. 
We will be studying causal time delay of 
low-energy excitations added to a macroscopic string representing 
quark or meson \cite{reyyee1, maldacenawilson}. 
We will find an excellent agreement between anti-de Sitter supergravity
and super Yang-Mills theory results. 
In section 4, we will study brane-wave transmission via Thomson scattering,
but now in anti-de Sitter spacetime. In appendix A and B, we will explain
technical details of covariant Green-Schwarz action for open string stretched
between the two D3-branes, in particular, consistent gauge-fixing of 
$\kappa$-symmetry. We will also derive a low-energy effective Lagrangian for
bilocal W-boson, for both flat and anti-de Sitter spacetimes. 

\section{Causality and Locality in Flat Spacetime}
Before dwelling into anti-de Sitter
spacetime, we will first study causality and locality in flat spacetime. 
To make a 
direct comparison with anti-de Sitter spacetime later, as a probe, we will use 
a system consisting of two D3-branes and a open F- or D-string stretched 
between them.

Consider, in Type IIB string theory, two parallel D3-branes in flat spacetime
$X = {\bf R}^{9,1}$. The D3-branes are oriented along $0,1,2,3$ directions and 
are located along $4, \cdots, 9$ directions at $(r_1, 0, \cdots, 0)$ and 
$(r_2, 0, \cdots, 0)$, respectively. Low-energy dynamics on the D3-brane 
worldvolume is governed by $d=4, {\cal N}=4$ super Yang-Mills theory with 
gauge 
group $G = U(2)$. We will denote generators of $U(2)$ gauge group as 
$T_a$ $(a = 1,2,3,4)$, where $T_{1,2,3} = \sigma_{1,2,3}$ belong to the 
$SU(2)$ and $T_4 = id$ to the diagonal $U(1)$ subgroups. 
As the two D3-branes are 
separated by a distance $\Delta r = \vert r_1 - r_2 \vert$, the $SU(2)$
gauge group is spontaneously broken to $U(1)$ generated by $T_3$ Cartan
subalgebra.
Together with the diagonal $U(1)$, the gauge group on the two D3-branes, 
which we label by 1 and 2, is then given by  $H = U^{(1)}(1) \times 
U^{(2)}(1)$, generated by diagonal linear combinations, $(T_3 \pm T_4)/2$.
A massive W-boson (or dual magnetic monopole) associated with the spontaneous
symmetry breaking is represented by an open F-string (or D-string) stretched 
between the two D3-branes. The static mass of the W-boson and the dual 
magnetic monopole is given by 
\be
M_{\rm w} = T \Delta r , \qquad M_{\rm m} = {T \over g_{\rm st}} \Delta r.
\label{massrelation}
\ee
They are part of an isospin triplet 
under $G$ (or its dual gauge group) and hence, under $H$, carry electric
(or magnetic) charges $(Q, - Q)$ where $Q = \pm 1$ \footnote{Charge 
conjugation on the D3-brane worldvolume is generated by worldsheet parity 
reversal of the attached open strings.}. 

To investigate causality and locality of the flat Minkowski spacetime over the
distance scale $\Delta r$, one will need to excite slightly, say, the first
D3-brane and follow subsequent propagation of the excitation, which will
eventually arrive at and excite the second D3-brane. In the limit $g_{\rm st}
\rightarrow 0$, semi-classical dynamics of the Type IIB supergravity ought to
obey both locality and causality. For low-energy excitation, the leading order
process in this limit will be such that the excitation is transmitted from
the first D3-brane to the second through an open F- or D-string stretched
between them \footnote{Processes involving more than one open strings or
closed strings are at least ${\cal O}(g_{\rm st} )$. At weak coupling,
$g_{\rm st} \rightarrow 0$, their contribution is negligible.}.
When exciting brane-wave on the first D3-brane, there are two channels
available: massless Higgs or $U^{(1)}(1)$ gauge field waves. 
In what follows, we will consider exclusively adding a weak amplitude, 
plane-wave of the $U^{(1)}(1)$ gauge field on the first 
D3-brane \footnote{
The Higgs field interactions can be included analogously from that 
of gauge fields, as they are
related by underlying ${\cal N}=4$ supersymmetry. }.
The entire dynamical process is then viewed as the classic, Thomson 
scattering of the $U^{(1)}(1)$ radiation field off the open string.

The open string stretched between the two D3-branes is casually identified 
with a massive W-boson.
This sounds paradoxical since the open string is an extended object whose
characteristic scale is the string scale, $\ell_s$, while the massive
W-boson is a point-like object. An answer to this is 
well-known \cite{polchinski} ever since the advent of the D-branes. 
In this section, from the dynamical point of view, we will revisit this issue 
as it is 
intimately tied with understanding causality and locality of the ambient 
spacetime in which the D-branes are embedded. Specifically, we will study 
the Thomson scattering process in two opposite limits of the massive W-boson, 
first in a point-particle limit and second in a stretched open string limit. 
We will find shortly that, over the distance scale $\Delta r$, super Yang-Mills
theory will perceive nonlocality and acausality in the limit the W-boson 
may be treated as a point-particle. In order to recover locality and causality,
the W-boson ought to behave more like an extended, bilocal object.  
Crossover between the two limits takes place, as anticipated, at $\Delta r 
\approx {\cal O}(\ell_s)$, the minimum distance scale of the ambient spacetime 
probed by a F-string. Somewhat surprisingly, we will find that the
same conclusion holds for a massive magnetic monopole, viz. an open D-string
stretched between the two D3-branes.

\subsection{Scattering of Flat Brane-Wave by Point-Like W-Boson} 
As is set up, a plane-wave of monochromatic radiation of $U^{(1)} (1)$ 
gauge field is incident on a free, massive W-boson of charge $Q$ and mass 
$M_{\rm W}$. 
The W-boson will be accelerated and emit radiation and, if the energy of the 
incident
radiation is low enough compared to the mass $M$, the emitted radiation will 
have the same frequency as the incident radiation. The whole process then
can be described by conventional Thomson scattering process, for which the 
W-boson is treated as a point-particle.
As the W-boson is charged
under $H$, the emitted radiation will consist of both $U^{(1)}(1)$ and
$U^{(2)}(1)$ gauge fields field. In terms of the D3-branes, this means that
both D3-branes will be excited by dipole oscillation of the stretched open 
string (interpreted as the massive W-boson) after the incident 
$U^{(1)}(1)$ radiation on the first D3-brane scatters off it, as 
depicted in figure 1.

\begin{figure}[htb]
   \vspace{0cm}
   \centerline{
\epsfig{file=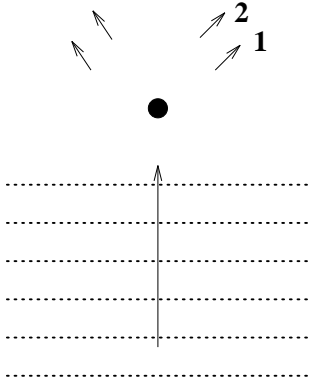,height=5cm,angle=0}
              }
\caption{Thompson scattering of brane wave off a W-boson. The incident 
$U^{(1)}(1)$ brane wave, whose wave front is denoted by dashed lines,
is scattered off a point-like massive W-boson at rest.
The scattered spherical brane waves of both $U^{(1)}(1)$ and $U^{(2)}(1)$
gauge groups, labelled as 1,2, are emitted instantly and simultaneously. }
\end{figure}

\subsubsection{Scattering Equation of Motion} 
We will study dynamics of a massive W-boson, located at ${\bf X}(t)$, 
interacting with gauge fields of the gauge group $H$. From the D3-brane
point of view, the W-boson is realized by an open F-string stretched between
the two D3-branes. Denote spatial position (on the two D3-branes) of
its endpoints as ${\bf X}_{1,2}(t)$. In the limit the W-boson is treated as
a point-particle, the position of the W-boson ${\bf X}(t)$ ought to coincide
with the two endpoint positions, ${\bf X}_{1,2} (t)$ all the time.
Consider an incident plane-wave radiation of the $U^{(1)}(1)$ gauge field, 
whose wave vector and 
frequency are denoted by $\bf k$ and $\omega$, respectively.
As the W-boson is charged with charge $(Q, -Q)$ under $H$, electric field 
of the incident radiation:
\be
{\cal E}({\bf x}, t) = {\bf E}^{(1)} (t) \, e^{ i {\bf k} \cdot {\bf x}}, 
\qquad {\bf E}^{(1)}(t) = {\bf E}_0^{(1)} e^{ - i \omega t},
\nonumber
\ee
according to the Lorentz force equation, 
will exert an {\sl instantaneous} acceleration to the massive W-boson:
\be
\ddot {\bf X}_1(t) = \ddot {\bf X}_2 (t) = {Q \over M_{\rm w}} 
{\bf E}^{(1)} (t) e^{ i {\bf k} \cdot {\bf x}}.
\label{eom}
\ee
Throughout this paper, we will consider the weak field-strength limit. 
In this case, the W-boson is in nonrelativistic motion and, at leading order,
the spatial dependence of the electric field on the right-hand side of 
Eq.(\ref{eom}) may be ignored.

The Eq.(\ref{eom}) shows that the two ends, even though separated by a distance
$\Delta r$, accelerates in an identical manner, viz. the 
stretched open F-string behaves like a rigid rod. 
It indicates that, should the open F-string be treated as a point-particle, 
super Yang-Mills theory on the D3-brane worldvolume would entail inevitably
nonlocality over the distance scale $\Delta r$ in the direction perpendicular 
to the D3-brane. This is of course as it should be, in order for the 
perpendicular direction to be interpreted as color isospin direction. 

\subsubsection{Transmission Rate }
To appreciate the significance of the Eq.(\ref{eom}), we will now calculate
the rate of energy transmission $T$ from the first D3-brane to the second 
through the stretched open string. 
From Eq.(\ref{eom}), one can determined the instantaneous power $P(t)$ 
radiated into a polarization state $\bf \varepsilon$ by the W-boson is
\footnote{For a weak field limit of the incident radiation, the W-boson
may be treated as heavy enough. In this case, spin and charge degrees 
of freedom decouple each other. We will henceforth treat the W-boson
as a spinless particle.}
\be
{d P \over d \Omega_2} = {Q^2 \over 4 \pi c^2} \vert {\bf \epsilon}^* \cdot
\ddot {\bf X}(t) \vert^2 .
\label{transmissiondef}
\ee
Averaging over a period $2 \pi / \omega$, during which 
the charged particle moves a negligible fraction of one wavelength,
\be
\Big\langle {d P \over d \Omega_2 } \Big\rangle 
= c {{{\bf E}_0^{(1)}}^2 \over 8 \pi} \left( {Q^2 \over M_{\rm w} 
c^2} \right)^2 
\vert \epsilon^* \cdot \hat {\bf E}_0^{(1)} \vert^2 .
\nonumber
\ee
The first factor $({{\bf E}_0^{(1)}}^2 / 8 \pi) c$ 
is nothing but the incident energy 
flux, viz. the time-averaged Poynting vector for the plane wave. 
Hence, differential transmission rate for an unpolarized incident 
radiation is given by
\be
\left( {d T \over d \Omega_2} \right)_{\rm classical}
= \left( {Q^2 \over M_{\rm w} c^2} \right)^2 \cdot
{1 \over 2} \left(1 + \cos^2 \theta \right),
\label{classical}
\ee
in which $\theta$ denotes the scattering angle in the laboratory frame. 
The total transmission rate $T$ obtained thereof is
\be
T_{\rm classical} = {8 \pi \over 3} \left( {Q^2 \over M_{\rm w} c^2} \right)^2.
\nonumber
\ee 
Characteristic feature of the classical Thomson scattering is that the 
cross-section is 
{\sl independent} of the frequency of the scattered radiation. 
The result Eq.(\ref{classical}) is valid only at low frequency limit,
$\omega \ll Mc^2$, for which the gauge field can be treated as a classical
wave. As the energy of the incident radiation becomes larger and, especially,
comparable to the W-boson mass, the scattering process should be treated 
quantum mechanically, viz. treating the radiation as photons. Treated in
the Coulomb gauge for which the transition matrix element is identical to 
the classial amplitude, the modification is from the phase space factor and
hence is purely kinematical. The result is
\be
\left( {d T \over d \Omega_2} \right)_{\rm quantum}
= \left( {Q^2 \over M_{\rm w} c^2} \right)^2 
\left( \omega' \over \omega \right)^2  \cdot {1 \over 2} (1 + \cos^2 \theta),
\nonumber
\ee
where the ratio of the outgoing to the incident frequency is given by
the well-known Compton formula:
\be
{\omega' \over \omega} = \left( 1 + 2 {\omega \over M_{\rm w} c^2} 
\sin^2 {\theta \over 2} \right)^{-1}.
\nonumber
\ee 
For low frequency limit, $\omega \ll M_{\rm w} c^2$, one easily obtains the 
transmission rate as
\be
T_{\rm quantum} 
= {8 \pi \over 3} \left( {Q^2 \over M_{\rm w} c^2} \right)^2 
\left( 1 - 2 {\omega \over M_{\rm w} c^2} + \cdots \right).
\label{pointparticle}
\ee

Had we have considered a point-like limit of the magnetic monopole 
dual to the W-boson, represented by an open D-string stretched between
the two D3-branes, the transmission rate would be essentially the  
same as above except that $M_w$ is to be replaced by the mass of the magnetic
monopole $M_m$ and that the charge Q is interpreted as the dual magnetic
charge \cite{baklee}. 

To recapitulate, in the limit the W-boson is point-like, viz. the stretched 
F-string moves rigidly, the transmission of radiation energy from the first 
D3-brane to the second through W-boson is {\sl instantaneous} and is a 
consequence of the standard field theory result, Eq.(\ref{eom}). Based on this
fact, we conclude that the D3-brane worldvolume dynamics, if treated in terms
of conventional super Yang-Mills theory and point-like W-bosons, would 
perceive nonlocality and acausality over a distance scale $\Delta r$ in the 
six-dimensional transverse space in $X$. From the super Yang-Mills theory 
point of view, this conclusion should be hardly surpring as the color isospin 
space is only an internal space and is not part of the four-dimensional 
spacetime (worldvolume of the D3-branes). 

\subsection{Scattering of Flat Brane-Wave by Charged Open String}
From the underlying string theory point of view, nonlocality and
acausality over the distance scale $\Delta r$ are extremely bizzare. 
The transverse to the D3-branes is six-dimensional subspace in $X$ and
string theory ought to exhibit locality and causality, at least over a
long distance limit. For example, energy transfer from the first D3-brane
located at $r=r_1$ to the second at $r=r_2$ through the open string between
them ought to take a causal time delay
\be
\Delta t \equiv {1 \over c} (r_2 - r_1) = {\Delta r \over c},
\label{timedelay}
\ee
as depicted in figure 2.

\begin{figure}[htb]
   \vspace{0cm}
   \centerline{
\epsfig{file=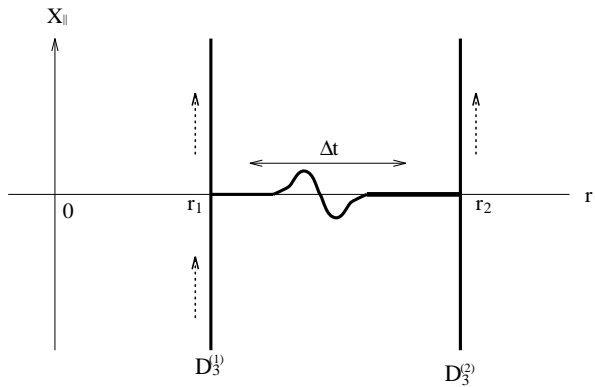,height=5cm,angle=0}
              }
\caption{ a stretched F- or D-string connecting two probe D3-branes
in flat spacetime background. vibration of the stretched string causes
retarded signal propagation between the two D3-branes only when the separation
is bigger than the string scale.}
\end{figure}

Resolution of the puzzle is quite simple. As nonlocality and acausality
has arisen from the point-particle limit of the W-boson, one expects that
locality and causality will become transparent in the opposite limit, 
viz. the W-boson (realized as an open string stretched between the two 
D3-branes) is treated as a full-pledged string. This turns out correct 
and means that one will now
need to study the Thomson scattering of the incident $U^{(1)}(1)$ radiation
off a string-like W-boson. In fact, in this section, 
we will find that the W-boson dynamics is drastically modified from that in 
the point-particle limit and that the modified dynamics is precisely what 
allows to restore locality and causality of $X$. 

\subsubsection{Scattering Equations of Motion}
Consider again the Thomson scattering of the $U^{(1)}(1)$ field radiation
off the W-boson,
but now in the limit the W-boson is treated as a full-fledged open 
string. Dynamics of the string is governed by Type IIB Green-Schwarz action, 
supplemented by an appropriate boundary action for the open string attached
to the D3-branes. In Nambu-Goto formulation,
which is sufficient for the Thomson scattering process, the action is 
constructed in Appendix A.    
After fixing the $\kappa$-symmetry in a gauge compatible with the boundary 
conditions and worldsheet reparametrization symmetries in static gauge,
whose steps are explained in detail in Appendix A, one finds that the
action for the string transverse coordinates ${\bf X}(r,t)$ is reduced 
to the form:
\bee
I_{\rm string}
&=& \int dt \left[ - M_{\rm w} +  
{T \over 2} \left( {1 \over c^2} \left( \partial_t {\bf X} \right)^2 
- \left( \partial_r {\bf X} \right)^2 \right) + \cdots \cdots \right]
\label{flataction} \\
&+& 
 \sum_{\rm I = 1, 2} \int dt \, Q_{\rm I}
\left[ A_0^{\rm (I)}({\bf r}, t) + {1 \over c}
\dot {\bf X}_{\rm I} \cdot {\bf A}^{\rm (I)}
( {\bf r}, t) \right]_{r = {\bf X}_{\rm I}(t)},
\qquad (Q_1 = - Q_2 = \pm 1).
\nonumber
\eee
The first term in the expansion represents the static mass of the string,
$T \Delta r$, which equals to the W-boson mass, $M_{\rm w}$ (see 
Eq.(\ref{massrelation})).
Consider, as posed above, Thomson scattering of a low-energy, monochromatic 
plane-wave of the 
$U^{(1)}(1)$ gauge field off the string endpoints. 
The boundary 
condition of the string coordinates ${\bf X}(t, r)$ is given by
\bee
T \partial_r {\bf X} (t, r)\Big\vert_{r = r_1} &=&  Q {\bf E}^{\rm (1)} (t) 
\nonumber \\
T \partial_r {\bf X} (t, r)\Big\vert_{r = r_2} &=& 0 \,\,\,.
\label{boundarycond}
\eee
According to the first boundary condition, 
the string endpoint ${\bf X}_1(t)$ (on the first D3-brane) 
will undergo a dipole oscillation and generate a pulse that will propagate 
subsequently along the string. The pulse may be decomposed into
spectral components
\be
{\bf X}(t, r) = \int {d \omega \over 2 \pi}
\left[ \,\,  {\bf a}(\omega) \,  e^{-i \omega(t - r/c)} \,\, +
\,\,
\tilde{\bf a} (\omega) e^{-i \omega (t + r/c)} \,\, \right].
\label{modeexpansion}
\ee
The spectral amplitudes ${\bf a}(\omega), \tilde{\bf a}(\omega)$ are 
determined uniquely by the Thomson scattering boundary condition 
Eq.(\ref{boundarycond}):
\be
{\bf a}(\omega) \,\, = \,\, \tilde{\bf a}^* (\omega) \,\, = \,\,
{Q {\bf E}_0^{(1)} \over 2 \omega T} 
\left( \cot \omega \Delta t - i \right).
\label{coefficients}
\ee
From Eqs.(\ref{modeexpansion}, \ref{coefficients}), 
one immediately obtains an equation of motion
for the string endpoint ${\bf X}_1(t)$ on the first D3-brane:
\be
M_{\rm w} \left( 
{\tanh \Delta t \partial_t 
\over \Delta t \partial_t} \right)
 \ddot {\bf X}_1 (t) \,\, = \,\, Q {\bf E}^{(1)} (t).
\label{leftend}
\ee
One first finds that the inertia mass of the open string equals to 
the static mass, $M_{\rm w}$, as is dictated by the 
underlying Lorentz invariance. 
Compared to the point-particle limit of the W-boson, Eq.(\ref{eom}), 
equation of motion of the endpoint ${\bf X}_1(t)$ is changed 
to the one involving a kernel of infinite-order time derivatives. 
As will be shown in section 
2.3, the extra kernel, which has
originated from taking into account of the string oscillations,  
is what enables low-energy dynamics of the D3-brane worldvolume to 
reproduce locality and causality of the ambient spacetime $X$. For now, 
note that, due to the kernel,  one may interpret the endpoint 
${\bf X}_1(t)$ to behave {\sl heavier} than the point-like 
W-boson mass $M_{\rm w}$. 

What about the dynamics of the second endpoint ${\bf X}_2(t)$ attached to the 
second D3-brane? For point-particle limit of the W-boson, we have seen that
the two endpoints behaves identically, see Eq.(\ref{eom}).
What one now finds, however, is that the dynamics of the second endpoint is
governed by a different equation of motion:
\be
M_{\rm w} 
\left( {\sinh \Delta t \partial_t 
\over \Delta t \partial_t} \right)
\ddot {\bf X}_2 (t) \,\, = \,\, Q {\bf E}^{(1)}(t).
\label{rightend}
\ee
Comparison with Eq.(\ref{leftend}) shows that the infinite-order kernel
associated with the second endpoint is different from that of the first one.
In particular, the kernel lets us to interpret that the seocond endpoint
${\bf X}_2(t)$ behaves {\sl lighter} than the point-like W-boson mass
$M_{\rm w}$. 

The Eqs.(\ref{leftend}, \ref{rightend}) constitute defining equations for the
Thompson scattering off the open F-string. Compared to the Thomson
scattering off the massive W-boson, new effects have entered through 
D3-brane-separation-dependent (see Eq.(\ref{timedelay})), infinite-order
kernels. As the two endpoints obey different equations of motion, at any
moment, one would expect that ${\bf X}_1(t) \ne {\bf X}_2(t)$. 
Therefore, from the super Yang-Mills theory perspective, it suggests to view
the W-boson more naturally as a sort of an extended, bi-local object. 

\subsubsection{Transmission Rate Across Open String}
Following the same steps as in section 2.1.2, 
it is straightforward to calculate the transmission rate $T$ (defined earlier
in section 2.1.2) from the
new equation of motion, Eq.(\ref{rightend}) of the string endpoint on the
second D3-brane. 

Actually, one will need to take into account of another important corrections 
to the transmission rate: virtual W-boson pair and radiation reactive force
effects. 
They are entirely of field-theoretic origin and have nothing to do with 
string oscillations. 
Because of these effects, in field theory treatment, the W-boson might be
viewed roughly as a particle with an effective size set by the W-boson 
Compton wavelength. Heuristically, the effects may be understood as follows
\cite{bakmin}.
Low-energy dynamics of a single W-boson may be described by taking a 
non-relativistic limit of the W-boson gauge field $W_\mu$:
\bee
W_0 &=& {1 \over \sqrt{ 2 M_{\rm w}} }
{i \over M_{\rm w} c} e^{- i M_{\rm w} c^2 t} \,
{\bf D} \cdot {\bf \Phi} ({\bf r}, t)
\nonumber \\
{\bf W} &=& {1 \over \sqrt{ 2 M_{\rm w}} }
e^{ - i M_{\rm w} c^2 t} {\bf \Phi} ({\bf r}, t),
\eee
in which charge-conjugate, anti-particle part is suppressed. 
Expanding the super Yang-Mills Lagrangian in the non-relativistic limit,
one finds
\be
{\cal L}_{\rm SYM}
= -{1 \over 4} F^{\mu \nu} F_{\mu \nu} 
+ i {\bf \Phi}^\dagger \cdot D_t {\bf \Phi}
+ {1 \over 2 M_{\rm w}} {\bf \Phi}^\dagger \cdot {\bf D}^2 {\bf \Phi}
- i {Q \over M_{\rm w} c} {\bf B} \cdot {\bf \Phi}^\dagger \times
{\bf \Phi} + \cdots. 
\ee
This can be reduced further to a single particle Lagrangian of the W-boson:
\bee
L_{\rm w} &=& {1 \over 2} m_{\rm w} \dot {\bf X} \cdot \dot {\bf X}
+ Q \left( A_0 + {1 \over c} \dot {\bf X} \cdot {\bf A} \right) + {\bf S}
\cdot {\bf B} + \cdots
\nonumber \\
&-& \int d{\bf r} \left( {1 \over 4} F^{\mu \nu} F_{\mu \nu} \right).
\eee
Here, parts containing classical spin dynamics and interaction with Higgs
field are suppressed for brevity. Following Abraham-Lorentz-type approach
\cite{jackson}, it is straightforward to derive an effective equation of 
motion that takes into account of radiative reaction force. One obtains
\be
M_{\rm w} \left( 1  - {Q^2 \over 4 \pi} {\partial_t \over M_{\rm w}}
+ \cdots \right) \ddot {\bf X}(t)
= Q {\bf E}_0(t).
\ee
If one takes into account of virtual W-boson pair corrections, one also
obtains a term similar to the second on the left-hand side. At low frequency,
the second term is suppressed compared to the first by 
${\cal O} (\omega / M_{\rm w})$ and hence, according to 
Eq.(\ref{transmissiondef}), gives rise to ${\cal O}
(\omega^2 / M^2_{\rm w})$ correction to the transmission rate. 

Taking into account of the above effects, one obtains the transmission rate 
for F-string as:
\be
T_{\rm F-string} = {2 \pi \over 3} \left( {Q^2 \over M_{\rm w} c^2} \right)^2 
\left( 1 - 2 {\omega \over M_{\rm w} c^2} + \cdots \right)
\cdot
\left( 1 - \left( {\omega \over M_{\rm w} c^2} \right)^2 + \cdots \right)
\cdot 
\left[ {\omega \Delta t \over \sin \omega \Delta t} \right]^2.
\label{flatfstringtransm}
\ee
The first two brackets represent the result derived from the limit the
W-boson is treated as a point-particle. The third bracket is the correction
due to virtual W-boson and radiation reactive force effect estimated above. 
The last bracket comes from the string oscillation effects, viz.
from the fact that the massive W-boson is actually a stretched open string. 

It is instructive to consider the transmission rate for, instead of a 
F-string, a D-string is stretched between the two D3-branes.
From super Yang-Mills point of view, the open D-string corresponds to 
a magnetic monopole dual to W-boson. In this case, scattering of the brane
wave takes place from interaction of the open D-string with magnetic 
component of the radiation field. The corresponding transmission rate is
obtained straighforwardly: 
\be
T_{\rm D-string} = {2 \pi \over 3} \left( {Q^2 \over M_{\rm m} c^2} 
\right)^2
\left( 1 - 2 {\omega \over  M_{\rm m} c^2} + \cdots \right)
\cdot
\left( 1 - \left( {\omega \over M_{\rm w} c^2} \right)^2 + \cdots \right)
\cdot
\left[ { \omega \Delta t \over \sin \omega
\Delta t } \right]^2,
\label{flatdstringtransm}
\ee
where $Q$ should now be interpreted as magnetic charge of the dual W-boson.
One again finds that, apart from the kinematical, Compton scattering 
correction, corrections of ${\cal O}(\omega^2/M_{\rm w}^2)$ arise from 
the classical finite-size of the magnetic monopole, set by the Compton
wavelength of the W-boson, and radiation reactive force effects \cite{bakmin}. 

We emphasize that, for both W-boson and its dual magnetic monopole, the
correction due to finite-size and radiation reactive force effects is set
by the W-boson mass, $M_{\rm w}$. Given that it is the only low-energy 
threshold scale in the system at weak coupling limit, 
the fact that the correction
for both W-boson and magnetic monopole is governed by the same low-energy 
scale should not be surprising \footnote{ 
Incidentally, the correction also entails 
an interesting deviation from S-duality of the $d=4, {\cal N}=4$ super 
Yang-Mills theory \cite{bakmin}.}. 
\subsection{Locality and Causality in Flat Spacetime}
 
Having analyzed transmission of brane waves across open string at 
two different limits, we will now study causality and locality of ambient
spacetime as seen by super Yang-Mills theory.  

\subsubsection{Retarded Dynamics of String Endpoints}
We have already seen from Eq.(\ref{eom}) that,  
in the limit the massive W-boson behaves like a 
point particle, the string endpoint on the second D3-brane, despite
being separated by a distance $\Delta r$, 
responds to the incident brane wave localized on the first D3-brane 
instantaneously.
As such, scattered brane wave will come out of both isospin components of 
$T_3$ (labelling the two D3-branes) simultaneously, with no apparent causal 
time-delay between the two isospin components. See figure 1. 
Indeed, the transmission rate, Eq.(\ref{pointparticle}), is exactly the same 
as the classic Thomson scattering cross-section. As a result, 
one will conclude that any super Yang-Mills 
theory with pointlike W-bosons will not be able to encode causality and 
locality of ambient spacetime.

In the opposite limit where the W-boson is a full-fledged oscillating string,
Eq.(\ref{modeexpansion}) shows that a pulse propagating along the string
is described by characteristic functions of the form $f(t \pm r /c )$. Clearly,
it will take a causal time delay $\Delta t$, Eq.(\ref{timedelay}), 
for the pulse, induced by the incident brane wave on the first D3-brane,
to reach the second D3-brane. To appreciate the causal time-delay more
transparently, let us rewrite the equation of motion of second string endpoint,
Eq.(\ref{rightend}), into an integral equation form:
\be
M_{\rm w} \ddot {\bf X}_2 (t)
= \int dt'  \, \, \Theta (t - t')
\bigg\langle t \bigg\vert {\Delta t \partial_t \over \sinh \Delta t \partial_t}\bigg\vert t' \bigg\rangle \cdot Q {\bf E}^{(1)}(t'),
\label{integraleqn}
\ee
where the Heaviside step function follows from convergence requirement, 
as is usual in any nonrelativistic dynamics. The integral kernel shows 
manifestly that the time-delay between Note that the integral kernel
has infinitely many poles but no zeros.

The equation of motion, Eq.(\ref{rightend}), can be put into an equivalent
to Eq.(\ref{integraleqn}), but even
more suggestive form by observing the fact that the infinite-order kernel
is in fact a {\sl difference operator}: 
\be
{\bf X}_2 (t + \Delta t) - {\bf X}_2 (t - \Delta t)
=  2\Delta t \int^t dt' \, Q {\bf E}^{(1)} (t'),
\ee
displaying manifestly that it takes causal time-delay $\Delta t$ for the
brane wave on the first D3-brane, ${\bf E}^{(1)}$, to arrive at the second.
The fact that the time-delay is exactly $\Delta t$ as estimated from the
geometric distance between the two D3-branes, Eq.(\ref{timedelay}) also
implies that the boundary interactions between F-string and D3-branes 
are local. 

While encoding causality and locality quite successfully, the massive
W-boson in this limit is far from the point-like object one casually take
in Yang-Mills theory. Rather, the two endpoints ${\bf X}_{1,2}(t)$ of the
open string evolves independently. Hence, projected on D3-brane worldvolume,
the W-boson is now replaced by a sort of extended, bilocal object whose 
dynamics is captured by the equations of motion, Eqs.(\ref{leftend},
\ref{rightend}). 
 
\subsubsection{Physical Thresholds: `Winding' versus `Momentum'}
As is evident from the transmission rates, Eqs.(\ref{flatfstringtransm},
\ref{flatdstringtransm}), frequency-dependent corrections are characterized
by two distinct physical threshold scales inherent to the problem. The first
is set by the W-boson mass, $M_{\rm w} = T \Delta r$, defining a physical
threshold (of field theoretic origin) above which description in terms of
abelian gauge theory with gauge group $H$ breaks down. 
The second is set by energy gap of string oscillation, $1/\Delta t
= c/\Delta r$, above which a point-particle description of the W-boson breaks
down. Being proportional to $\Delta r$ and inversely to $\Delta r$, 
respectively, we shall be referring the first threshold as `winding threshold' 
$\Lambda_{\rm w}$ and the second as `momentum threshold', $\Lambda_{\rm m}$:
\be
\Lambda_{\rm w} = M_{\rm w} = T \Delta r, \qquad \quad
\Lambda_{\rm m} = {1 \over \Delta t} = {\hbar c \over \Delta r}.
\ee
As there are two distinct physical thresholds, depending on which one sets
the lower scale, low-energy dynamics of the super Yang-Mills theory
will become quite
different. As the winding and momentum thresholds depend on the 
spatial separation $\Delta r$ inversely, the two will competes each other
as $\Delta r$ is varied. The competition may be summarized into a form
of string uncertainty relation ($T \equiv 1/\ell_s^2$)
\be
\Lambda_{\rm w} \Lambda_{\rm m} \quad \approx 
\quad {\cal O}\left( {\hbar c \over \ell_s^2} \right).
\ee
Note that the right-hand-side, a combination of fundamental constants
only, depends crucially on the string scale, $\ell_s$. It implies that
any effects related to causal retardation, which arises from oscillation
threshold corrections, will inevitably depends on it, as $\partial / 
\Lambda_{\rm m} \approx \ell_s^2 M_{\rm w} \partial$. This explains the
observation of Douglas and Taylor \cite{douglastaylor} on $F^4$ interaction 
between brane waves with nonzero momentum. As $M_{\rm w} \propto \Delta r
\propto \Delta t$, one again finds that successive power corrections in 
oscillation threshold are nothing but the retardation time expansion of
Eq.(\ref{integraleqn}).

Hence, if $\Lambda_{\rm w} \ll \Lambda_{\rm m}$, threshold effects 
to low-energy processes dominated by the `winding threshold'. As string 
oscillation excitations are completely irrelevant, 
the stretched open string behaves literally like a rigid-rod, to which the 
D3-branes serve as a pair of guiderail. Behaving as a rigid
body, the open string may be treated as a point-particle, massive
W-boson. As the condition
\be
\Lambda_{\rm w}  \ll \Lambda_{\rm m} \quad
\rightarrow \quad \Delta r \ll \ell_s,
\label{condition}
\ee
one find that the limit in which the D3-brane worldvolume dynamics is best
approximated by super Yang-Mills theory, either in conventional (renormalizable
dimension-four operators) or Dirac-Born-Infeld form, is when the D3-brane
separation is over a sub-stringy scale, reproducing well-known result 
\cite{polchinski}.
  
Exactly the same condition applies if the stretched F-string is replaced
by a D-string. An open D-string stretched between two D3-branes is casually
identified with a magnetic monopole dual to the massive W-boson. 
Its classical size is set by the Compton wavelength of the massive W-boson 
$M_{\rm w} = T \Delta r$ \footnote{At weak coupling,
classical size of the stretched D-string $T \Delta r$ is much larger than
the Compton wavelength of the monopole $T \Delta r / g_{\rm st}$ and hence
a semi-classical treatment of the monopole is justifiable.} and defines the
field theoretic, winding threshold scale $\Lambda_{\rm w}$. 
Requirement that this scale is much less than the oscillation threshold scale
$\Lambda_{\rm m}$, which is the same for both F- and D-string, again set the
condition, Eq.(\ref{condition}). 

\section{Dynamical Aspect of Energy-Size Relation}
An important feature of the AdS/CFT correspondence is holographic
relation between the radial position, `energy scale', 
in anti-de Sitter spacetime
and `size' in super Yang-Mills theory. As this relation will be playing a 
crucial role for foregoing discusssions, we will be studying 
first dynamical aspect of the relation.  
More specifically, for heavy quark and meson studied in \cite{reyyee1, 
maldacenawilson}, we will estimate causal time-delay for a pulse generated
by a quark either along the string in anti-de Sitter spacetime or 
in Minkowski spacetime, using the harmonic fluctuation Lagrangian
of the string studied previously \cite{reyyee0, reyyee1, reyyee2}. 
We will find that, after utilizing the holographic 
energy-size relation, the two estimates agree perfectly with each other. 

A charged probe necessarily accompany long-range Coulomb field,
so it will be necessary to control the field configuration in the infrared. 
A convenient way is to move into the Coulomb branch, viz. separating some 
of the D3-branes and then study a state charged under the gauge group of 
the separated D3-branes. The $N$ coincident D3-branes sitting at the 
conformal point then generates the anti-de Sitter spacetime as a 
near-horizon geometry
\cite{maldacenaproposal} 
\be
ds^2 = \ell_s^2 \left[ \,
{U^2 \over g_{\rm eff}} \left( -{1 \over c^2} 
dt^2 + d {\bf x}_{\parallel}^2 \right)
+ g_{\rm eff} {dU^2 \over U^2} + g_{\rm eff} d\Omega_5^2 \, 
\right], \qquad \quad (U \equiv T r, \,\, T = 1/\ell^2_s).
\label{adsmetric}
\ee
In this coordinate, the energy-scale holographic relation asserts that
the scale position $U$ in the anti-de Sitter spacetime is related 
directly to the scale size $R$ in the super Yang-Mills theory:
\be
R \quad = \quad {g_{\rm eff} \over U}.
\label{uvir}
\ee
This relation has been first derived from evaluation of the Wilson 
loop \cite{reyyee1, maldacenawilson} and later extended to more
general contexts \cite{susskindwitten}.

\subsection{Holography of Heavy Quark Dynamics}
Consider a heavy quark of large-N super Yang-Mills theory at zero temperature.
We will separate a single D3-brane to the Coulomb branch and consider a 
quark coupled to it. Denote the location of the displaced probe D3-brane
as $U_*$. 
Then, a heavy quark (monopole) of the super Yang-Mills theory is realized
by a macroscopic F- (D-) string ending on the displaced D3-brane at 
$U = U_*$ and stretched outward along 
the $U$-direction.  Dynamics of the macroscopic string may be studied using 
Nambu-Goto action in the anti-de Sitter spacetime, Eq.(\ref{adsmetric}). 
Taking static gauge $(\tau, \sigma) = (t, U)$ and expanding string 
transverse
coordinates $({\bf X}_{\parallel}, \Omega_5)$ up to harmonic fluctuations, 
one finds \cite{reyyee1}
\be
L_{\rm NG} = - \int_{U_*}^\infty d U + \int_{U_*}^\infty dU {\cal L}^{(2)} 
 + L_{\rm boundary}
\label{nglag}
\ee
where
\be
{\cal L}^{(2)}
= 
{1 \over 2} \left[ \left( {1 \over c^2} {\dot {\bf X}}_{\parallel}^2
- {U^4 \over g^2_{\rm eff}} {{\bf X}'_{\parallel}}^2 \right)
+ U^2 \left( {g^2_{\rm eff} \over U^4} {1\over c^2} \dot \Omega_5^2 
- {\Omega'_5}^2 \right) \right] + \cdots
\ee
and $L_{\rm boundary}$ specifies boundary conditions at $U = U_*$.

Let us begin with {\sl static properties} of the macroscopic string.
The first term in Eq.(\ref{nglag}) represent the {\sl static mass} of the 
string
\be
M_{\rm string} (U_*) = \int_{U_*}^{\Lambda_{\rm UV}} d U, 
\hskip3cm \Lambda_{UV} \rightarrow \infty.
\ee
The static mass diverges as the cutoff $\Lambda_{\rm UV}$ is removed.
As shown first in \cite{reyyee1, maldacenawilson}
and discussed further \cite{susskindwitten, polchinskipeet}, 
this is nothing but anti-de Sitter space manifestataion of 
divergent self-energy of a heavy quark. A physically meaningful, finite 
quantity would be a {\sl residual static mass} 
between strings ending on different locations of $U_*$, which we may define
as
\be
\Delta M_{\rm string} (U_*) \equiv \int_0^{U_*} dU \, {d M_{\rm string}
\over d U }
= U_*. 
\ee

\begin{figure}[htb]
   \vspace{0cm}
   \centerline{
\epsfig{file=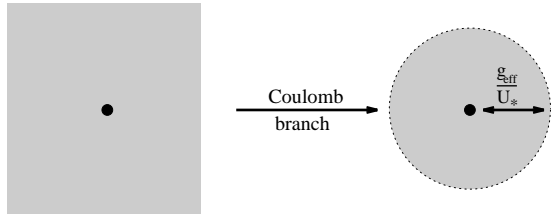,height=5cm,angle=0}
              }
\caption{
Coulomb branch view gauge field generated by a test quark (monopole) }
\end{figure}

Using the scale-size relation Eq.(\ref{uvir}), one then finds immediately
that $\Delta M_{\rm string} (U_*) = g_{\rm eff}/R_*$ corresponds to 
the Coulomb field energy measured at radius $R_*$. 
To understand this scale better, let us interpret the macroscopic string
from the super Yang-Mills theory side. 
At the origin of the Coulomb branch (viz. $U_* = 0$), the macroscopic string
creates not only a heavy quark (transforming as the defining representation of
$SU(N)$) itself but also long-range color Coulomb field around it, as
depicted in the left of Figure 3.
Away from the origin of the Coulomb branch, one might expects that the
long-range Coulomb field is cut-off at the Compton wavelength $U_*^{-1}$ of
the massive gauge bosons \footnote{Recall that the massive gauge bosons are
BPS particles}. However, as pointed out in \cite{reyyee1, maldacenawilson}, 
the strong `t Hooft coupling dynamics cuts off the long-range field at
a larger scale, $g_{\rm eff} U^{-1}_*$.
Thus, we find that $\Delta M_{\rm string}$ is nothing but the `field energy', 
the energy 
associated with Coulomb field outside radius $R_* = g_{\rm eff}/U_*$ in the
Yang-Mills theory. Summarizing, static property (such as static mass) of the
macroscopic string is determined by the quantity $\int d U$. 

\begin{figure}[htb]
   \vspace{0cm}
   \centerline{
\epsfig{file=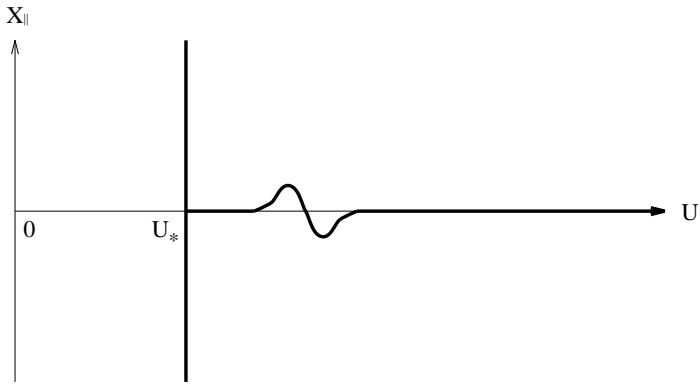,height=5cm,angle=0}
              }
\caption{string attached to a probe D3-brane in anti-de Sitter space}
\end{figure}

What about {\sl dynamical } properties? Consider exciting a low-energy pulse at
the outer end of the string, as depicted in Figure 4.. 
The pulse will then propagate at the speed of
light (which has been explicitly checked in \cite{reyyee1}) along the string 
toward $U_*$.  According to the energy-size relation, 
in super Yang-Mills theory, 
the pulse travelling along the string is mapped, in super Yang-Mills
theory, to a spherical brane wave generated at the position of 
the heavy quark and then propagating outward.
For the spherical wave to propagate over a radial size $R_*$, 
it will take Yang-Mills time:
\be
(\Delta t)_{\rm YM}
= 
{1 \over c} R_*.
\label{ymtime}
\ee

If causality is implied by the holographic energy-size relation, 
then the time-delay for the pulse propagation along the string should be 
the same as that for the spherical brane wave propagating outward. 
Let us check this explicitly.  
The string pulse propagates along the string, viz. a null geodesic in 
$(U, t)$ subspace:
\be
{U^2 \over g_{\rm eff} } \Delta t^2 = {g_{\rm eff} \over U^2}
\Delta U^2.
\ee
Thus, measured from anti-de Sitter spacetime in Yang-Mills time unit, 
causal time-delay for the pulse to travel from $U  = \infty$ to $U = U_*$ 
along the string is given by
\be
\left( \Delta t \right)_{\rm AdS} = {1 \over c} g_{\rm eff}
\int_{U_*}^\infty {dU \over U^2} = {1 \over c} {g_{\rm eff} \over U_*}. 
\label{adstime}
\ee
Remarkably, the holographic energy-size relation, Eq.(\ref{uvir}),  permits 
the causal time delay in the anti-de Sitter spacetime agrees exactly with
that along the boundary!
 
\subsection{Holography of Heavy Quark Dynamics at Finite Temperature}
Consider next the super Yang-Mills theory at nonzero temperature. 
The near-horizon geometry of near extremal D3-branes is described by
the anti-de Sitter-Schwarzschild spacetime, whose $(t, U)$ subspace is
given by:
\be
ds^2 =
\ell_s^2 \left[ \,
- {1 \over c^2} \left( 1 - {U_0^4 \over U^4} \right) {U^2 \over g_{\rm eff}}
dt^2 + {g_{\rm eff}  \over U^2} \left( 1 - {U_0^4 \over U^4} \right)^{-1}
dU^2 \, \right].
\ee
From Nambu-Goto Lagrangian, one again finds that the static mass of the string
is given by
\be
M = \int_{U_0}^\infty d U,
\ee
the same as the value at zero temperature. 

Let us introduce a heavy quark to the Yang-Mills theory at finite
temperature. The quark is described by a macroscopic F-string stretched 
along $U$-direction \cite{reyyee2, yank1}, as depicted in Figure 5.
In the present case, the string terminates at the Schwarzschild horizon,
$U = U_0$, where it will dissipate the F-string flux into the inner 
horizon region.
 
\begin{figure}[htb]
   \vspace{0cm}
   \centerline{
\epsfig{file=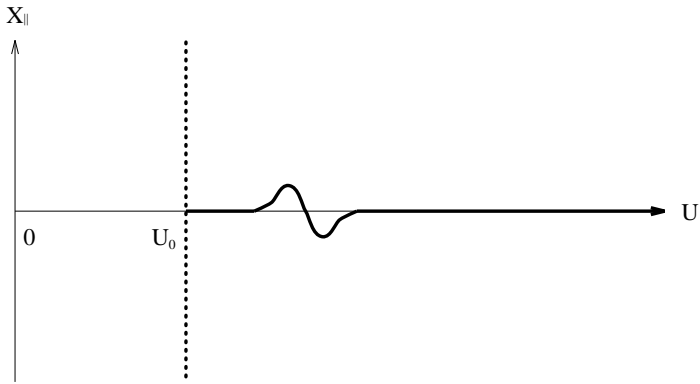,height=5cm,angle=0}
              }
\caption{ Pulse propagation along semi-infinte string in anti-de Sitter 
Schwarzschild spacetime.}
\end{figure}
 
This phenomenon has a direct counterpart in super Yang-Mills theory.
At zero temperature, Coulomb field of the static quark extends everywhere,
as depicted at the left in Figure 4. At finite temperature, however, the
Coulomb field becomes Debye-screened and is expected to extend to the scale
of Debye wavelength $U_0^{-1} = (g_{\rm eff} T)^{-1}$. Using the holographic
energy-size relation, one finds that the Coulomb field is cut off
at a larger scale $g_{\rm eff} U_0^{-1}$, again due to strong `t Hooft
coupling dynamics. See Figure 6.
 
\begin{figure}[htb]
   \vspace{0cm}
   \centerline{ \epsfig{file=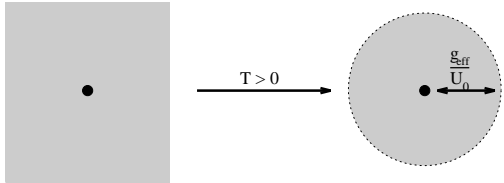,height=5cm,angle=0} }
   \caption{ aerial view of an isolated quark in super Yang-Mills 
                   theory at finite temperature}
\end{figure}
 
In fact, by demanding causality, one can understand the fact that string
must end at $U=U_0$. Consider a
weak pulse injected at $U = \infty$ and propagating inward. The time-delay
for the pulse to reach the position $U ( U_0 \le U \le \infty)$ is easily
estimated:
\be
(\Delta t)_{\rm AdS}
= {g_{\rm eff} \over U_0}
\int_0^{U_0/U} { dx \over 1 - x^4} .
\label{temptime}
\ee
The time-delay diverges logarithmically as $U \rightarrow U_0$. 
So, evolved in terms of Yang-Mills theory time, which is an asymptotic 
observer's time in the anti-de Sitter spacetime, the super
Yang-Mills theory covers spacetime region only outside the horizon.
If a different global time is chosen as an evolution operator, then
the string might be able to enter inside the horizon. In this case,
howeve, the string would not be in a static configuration \cite{banksetal}

At finite temperature, 
the scale $R_0 = g_{\rm eff} U_0^{-1}$ corresponds to a {\sl physical
threshold} scale in Yang-Mills theory. At weak `t Hooft coupling limit, 
Matsubara mass scale associated with dimensionally reduced fermions
(satisfying anti-periodic boundary condition around Euclidean time)
is ${\cal O} (T)$, which equals to $U_0 / g_{\rm eff}$.
If this observation somehow holds also for super Yang-Mills theory
at zero temperature, then we are prompted to expect that the scale
$R_* \equiv g_{\rm eff} U_*^{-1}$ actually corresponds to a new {\sl
physical threshold} scale. Recall that this scale is not a threshold
of an elementary excitations as  the Compton wavelength of massive 
gauge bosons $U^{-1}_*$. Nevertheless, the fact that the scale 
$R_*$ is $g_{\rm eff}$ times Compton wavelength of massive gauge bosons 
is tantalizingly similar to the observation of Shenker
made in a different context \cite{shenker}.

\subsection{Holography of Heavy Meson Dynamics}

\begin{figure}[htb]
   \vspace{0cm}
   \centerline{
\epsfig{file=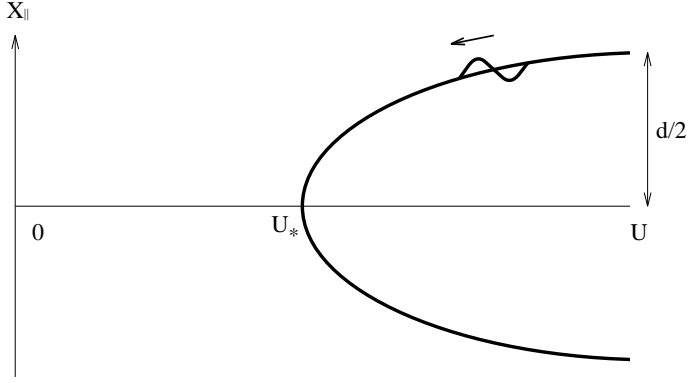,height=5cm,angle=0}
              }
\caption{A pulse propagatiing along the Wilson loop in anti-de Sitter 
spacetime.} 
\end{figure}

A more intricate bulk probe is the macroscopic string configuration 
corresponding to the heavy meson~\cite{reyyee1, maldacenawilson}. In the static
gauge, the string configuration is specified by a first-integral of motion:
\be
G^2 U'^2 + G = {g^2_{\rm eff} \over U_*^4} .
\label{1stintegral}
\ee
Again, consider a pulse propagating along the string, $U = U(\sigma)$.  
The time delay for the pulse to travel between the quark and anti-quark
(both `located' at $U = \infty$) along the string is again calculated 
straightforwardly by solving string equation of motion along the null 
geodesic:
\be
\sqrt{G} dU^2 = {1 \over \sqrt G} \left( -dt^2 + d \sigma^2 \right) \, .
\ee
This yields 
\be
(\Delta t)_{\rm AdS} = \int_{-d/2}^{+d/2} d \sigma \sqrt{ G U'^2 + 1}.
\ee
Using Eq.(\ref{1stintegral}), one finds immediately that
\bee
(\Delta t)_{\rm AdS} &=& 2 {g_{\rm eff} \over U_*^2} \int_{U_*}^\infty dU  
{1 \over \sqrt{ U^4/U^4_* - 1}} \nonumber \\
&=& {2 \sqrt{\pi} \Gamma(5/4) \over \Gamma(3/4)} {g_{\rm eff} 
\over U_*}.
\eee
From the super Yang-Mills theory side, natural scale of the meson is
the inter-quark distance $d$. Hence, one expects that
\be
\left( \Delta t \right)_{\rm YM} = A {1 \over c} d,
\ee
where $A$ is a ${\cal O}(1)$ numerical coefficient, which should reflect
the details of the dipole field configuration produced by the quark-antiquark
pair at strong `t Hooft coupling limit. For mesons, precise form of the 
holographic energy-size relation has been derived \cite
{reyyee1, maldacenawilson}: 
\be
{d \over 2} = {{\sqrt \pi} \Gamma(3/4) \over \Gamma(1/4)}
{g_{\rm eff} \over U_*},
\ee
Hence, equating $\Delta t$ of anti-de Sitter spacetime with that of
Yang-Mills theory, one finds that 
\be
A = {\Gamma(5/4) \Gamma(1/4) \over \Gamma^2(3/4)} = 2.188...
\ee

\begin{figure}[htb]
   \vspace{0cm}
   \centerline{
\epsfig{file=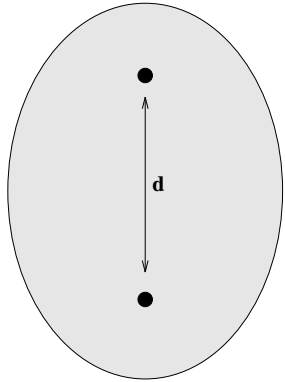,height=5cm,angle=0}
              }
\caption{ Aerial view of the quark anti-quark pair and dipole field 
configuration produced thereof.}
\end{figure}

The numerical factor, $A = 2.188...$, associated with the dipole
electric field configuration of the quark antiquark pair as depicted
in figure 8, may be
viewed as a prediction for the strong `t Hooft coupling dynamics
from anti-de Sitter supergravity.

\section{Causality and Locality in Anti-de Sitter Spacetime}
Having understood dynamical aspect of the energy-size relation better, we 
will now revisit the Thomson scattering of brane waves by an open string 
and consequent implications to the issue of causality and
locality, but now all in anti-de Sitter spacetime.

We will be showing that `momentum' threshold of the open string stays always 
below the `winding' threshold everywhere in the anti-de Sitter spacetime. 
Moreover, quite importantly, both thresholds does not depend on the string
scale, $\ell_s$. Thus, in sharp contrast to the flat spacetime situation, 
we will be able to encode causality and locality of semi-classical supergravity
in anti-de Sitter spacetime in terms of large-$N$ super Yang-Mills theory,
at least in the strong `t Hooft coupling limit. The resulting super Yang-Mills
theory is not the standard one, however. At at generic point of the Coulomb 
branch moduli space\footnote{except possibly at the origin, on which we will 
be discussing later}, we will be finding that infinite tower of long 
supermultiplet operators should be included to be able to encode the bulk 
causality and locality. Once again, this conclusion is drawn from the fact
that W-bosons are more suitably described by a sort of extended, bilocal
objects. 

\subsection{D3-Branes in the Coulomb Branch}
The anti-de Sitter spacetime is generated as near-horizon geometry of
infinitely many $(N \rightarrow \infty)$ D3-branes when they become 
coincident one another. Therefore, to explore causality and locality of the 
anti-de Sitter spacetime, from the viewpoint of super Yang-Mills theory, one 
will need to examine low-energy dynamics near the origin of the Coulomb branch 
moduli space, where the Higgs expectation values are interpreted as positions 
of individual D3-branes in anti-de Sitter spacetime \cite{maldacenaproposal,
douglastaylor}. 

As in flat spacetime, we will be considering two D3-branes out of $N$ 
coincident oneses are located at positions $U_1, U_2$, respectively, and 
a macroscopic open string stretched between them. 
We will again excite the displaced probe D3-branes with low-energy brane
waves and study Thomson scattering off the open string. The configuration
corresponds to the following sequential symmetry breaking:
\be
SU(N) \rightarrow G = S\left[U(N-2) \times U(2) \right]
\rightarrow H = S \left[U(N-2) \times (U(1) \times U(1) ) \right].
\ee

From super
Yang-Mills theory perspectives, this appears quite a complicated problem,
as the excitation of, say, first D3-brane will reach to the second D3-brane
not only through the prescribed open string stretched between them 
but also indirectly through the coincident $(N-2)$ D3-branes, with which 
the two D3-branes interact. At the least, because the number of  
possible transmission channels through the `background' 
$(N-2)$ D3-branes is humongous, ${\cal O}(N^2)$, one might be tempted 
to conclude that, in the super Yang-Mills theory, brane wave transmission 
is not mediated through the massive W-bosons but by some sort of 
complicated large-$N$, strong `t Hooft coupling dynamics. 

This, however, shouldn't be a problem. The point is that interaction between
the probe Yang-Mills theory with gauge group $U(2)$ or $U(1) \times U(1)$ and
that of conformal sector with gauge group $SU(N-2)$ are controlled by the
mass scales, $U_{1,2}$. Taking $\Delta U = \vert U_1 - U_2 \vert$ 
much lighter and $g_{\rm st}$ sufficiently small, dynamics of the two probe
D3-brane sector can be studied in a controllable manner. Thus, starting from
the super Yang-Mills theory with gauge group $G$, we will be integrating out 
over the Higgs branch, viz. open strings transforming in 
$({\bf N-2}, \overline{\bf 2})$ and its complex-conjugate representations. 
In the $g_{\rm st} \rightarrow 0$, $N \rightarrow \infty$ and
large `t Hooft coupling limit, the integration amounts to resummation of 
large-$N$ Feynman diagrams. As have argued in \cite{maldacenaproposal} 
(see also discussions in \cite{douglastaylor}), the result is to produce 
anti-de Sitter spacetime, inside which the two displaced probe D3-branes are 
localized.

Intuitively, the above assertion may be understood by observing that 
variance of positions of the $(N-2)$ coincident D3-branes and two displaced 
D3-branes are of order $(g_{\rm st} N)^{1/4}$ and $(g_{\rm st})^{1/4}$, 
respectively.
That is, in the large-N limit, position of the two probe D3-branes is
sharply localized inside the anti-de Sitter spacetime. It should also become 
clear that, even after adding a stretched open string between the 
two displaced D3-branes (associated with the symmetry breaking $G \rightarrow
H$), integration over the Higgs branch yields essentially the same result, 
viz. dynamics of both the two probe D3-branes and the stretched open string 
between them will perceive the ambient background as anti-de Sitter spacetime. 

Henceforth, at $g_{\rm st} \rightarrow 0$, $N \rightarrow \infty$ and 
strong `t Hooft coupling regime, we will be able to study causality and 
locality of the anti-de Sitter spacetime in a controllable manner by 
focusing only to low-energy excitations of massive W-bosons in `quantum
corrected' super 
Yang-Mills theory with symmetry breaking pattern $U(2) \rightarrow U(1) 
\times U(1)$. 

\subsection{Scattering of Brane Wave by Charged Open String}
Dynamics of the open F-string stretched between the two D3-branes is governed 
by the Type IIB Green-Schwarz action in the anti-de Sitter spacetime. In
Appendix B, for Nambu-Goto form of the action, details of gauge fixing of 
local $\kappa$- and reparametrization symmetries are explained. 
It turns out the gauge-fixing as well as analysis of string dynamics thereof
become enormously simplified once one adopts
a new coordinate, $R \equiv g_{\rm eff} / U$. According to the energy-size
holography relation, Eq.(\ref{uvir}), the new coordinate $R$ is seen 
nothing but the `size' variable of super Yang-Mills theory. 
It is interesting that, in analyzing a `bulk' process in 
anti-de Sitter spacetime, description in terms of a `boundary' variable 
turns out more suited and efficient. 
Indeed, in terms of the `size' coordinate $R$, the metric of the anti-de
Sitter spacetime is given in Poincar\'e coordinates:
\be
ds^2_{\rm AdS} = {g_{\rm eff} \over R^2} \Big[
- dt^2 + d {\bf x}^2_{\parallel} + d R^2  + R^2 d \Omega_5^2 \Big].
\nonumber \\
\ee
As elaborated in Appendix B, this coordinate
choice leads the Green-Schwarz action to an almost identical form, after
gauge fixing, to the one in flat spacetime, for both bosonic and fermions
parts.  Moreover, as has been observed in \cite{reyyee1}, harmonic dynamics
of the open string turns out considerably simplied in the Poincar\'e
coordinate system. After the gauge-fixing, the open string action is given
by
\bee
I_{\rm string}
&=& \int \! dt \,
\left[ - M_{\rm w} + {g_{\rm eff} \over 2}
\int_{R_1}^{R_2} {dR \over R^2} 
\left( {1 \over c^2} (\partial_t {\bf X})^2 - (\partial_R {\bf X})^2 \right)
+ \cdots \right] 
\label{actionads} \\
&+& \sum_{\rm I = 1,2} \int \! dt \, Q_{\rm I} \left[ 
A_0^{\rm (I)}({\bf r}, t) + {1 \over c} \dot {\bf X}_{\rm I} (t) \cdot 
{\bf A}^{\rm (I)}({\bf r}, t) \right]_{{\bf r} = {\bf X}_{\rm I}(t)}.
\eee

\begin{figure}[htb]
   \vspace{0cm}
   \centerline{
\epsfig{file=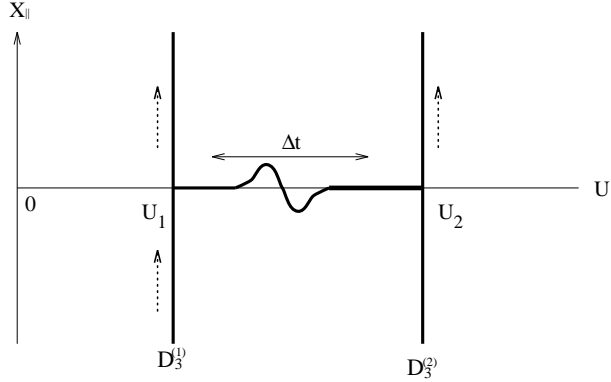,height=5cm,angle=0}
              }
\caption{a stretched F- or D-string connecting two probe D3-branes 
in the Coulomb branch. vibration of the stretched string causes retarded
radiation between the two D3-branes.}
\end{figure}

First, one notes that static mass of the string as measured in 
anti-de Sitter spacetime, the first term in Eq.(\ref{actionads}), 
$g_{\rm eff} \int dR /R^2$ equals to $(U_2 - U_1)$ in terms of the original
`energy' coordinate. Similar result holds also for an open 
D-string stretched between the two D3-branes.  Hence,
\be
M_{\rm w} = \Delta U,
\quad \qquad
M_{\rm m} = {1 \over g_{\rm st}} \Delta U.
\label{adsmass}
\ee
Recalling that $\Delta U = T \Delta r$ (cf. Eq.(\ref{adsmetric})), one
discovers that static mass of the string in anti-de Sitter spacetime is the
same as that in flat spacetime.
This is, from the super Yang-Mills theory perspective, as it should be. 
Both W-boson and dual magnetic monopole are BPS-saturated
states and hence their masses are not renormalized as one interpolates
from weak to strong `t Hooft coupling regime. 

From Thomson scattering process in flat spacetime, we have observed that the
static mass of the W-boson and the causal time-delay are governed by one
and the same length scale, $\Delta r$, the separation between the two 
D3-branes. As the W-boson is a BPS state, one might have guessed that it
will be the same in anti-de Sitter spacetime. After all, the distance between
the two D3-branes is the only relevant length scale inherent in the 
problem. This expectation, however, turns out not quite right. We will be
showing below 
that the causal time-delay is governed not by the scale $\Delta U$, but by
\be
\Delta \tau \equiv {1 \over c} \Delta R = {1 \over c} 
g_{\rm eff} \bigg\vert {1 \over U_1}  - {1 \over U_2} \bigg\vert.
\label{adstimedelay}
\ee
It is clear that $\Delta \tau$ has no simple functional relation to 
$\Delta U$, the scale that has set the W-boson mass for both flat and
anti-de Sitter spacetime and the energy variance that the spacetime 
uncertainty relation \cite{yoneya} utilizes. In fact, the reason why 
$\Delta (g_{\rm eff}/U)$, not $\Delta U$, should be the relevant variable
for the time-delay may be gleaned from the fact that $SO(4,2)$ invariant
line element between two points $( t, U, {\bf x})_{1,2}$ is given by    
\be
\Delta s^2_{\rm AdS} = U_1 U_2 \left[ - (t_1 - t_2)^2   
+ g_{\rm eff}^2 
\left({1 \over U_1} - {1 \over U_2} \right)^2 + ({\bf x}_1 - {\bf x}_2)^2
\right].
\ee
\subsubsection{Scattering Equation of Motion}
Consider excitation of a low-energy, monochromatic plane-wave of the
$U^{(1)}(1)$ gauge field on the first D3-brane. As in flat spacetime,
we will be studying Thomson scattering of the brane-wave off the open
string. We have argued above that, in so far as $g_{\rm st} = g_{\rm YM}^2 
\rightarrow 0$,
low-energy processes between the two probe D3-branes can be trusted. Actually,
the Thomson scattering process depends on the combination of $g_{\rm YM}
{\bf E}_0 / M_{\rm w}$. Thus, even in the situation where $g_{\rm YM}$ itself may not be
sufficiently small, the foregoing calculations will be completely controllable
by taking the brane waves sufficiently weak. 
 
In static gauge, the boundary condition of the transverse string coordinates
${\bf X}(t, R)$ is given by
\bee
{g_{\rm eff} \over R^2} \partial_R {\bf X}(t, R) \bigg\vert_{R = R_1}
\,\,\, &=& \,\,\, Q {\bf E}^{(1)}(t),
\nonumber \\ 
{g_{\rm eff} \over R^2} \partial_R {\bf X}(t, R) \bigg\vert_{R = R_2} 
\,\,\, &=& \,\,\, 0 \,\, .
\nonumber \\
\label{adsbc}
\eee
From Eq.(\ref{actionads}), one finds that a weak pulse propagating along the 
string is given by
\be
{\bf X}(t, R) = \int {d \omega \over 2 \pi}
\left[ \, {\bf a}(\omega) \, \left(1 - i \omega {R \over c} \right) 
e^{ - i \omega (t - R/c) }
+ \, \tilde {\bf a} (\omega) \, \left(1 + i \omega {R \over c} \right) 
e^{-i \omega (t + R/c) } 
\, \right].
\label{modeexpads}
\ee 
Again, the boundary condition, Eq.(\ref{adsbc}), fixes the spectral amplitudes 
${\bf a}(\omega), \, \tilde{\bf a}(\omega)$ uniquely:
\be
{\bf a}(\omega) = \tilde {\bf a}^* (\omega)
= \left( {R_1 / g_{\rm eff} \over \omega}\right) 
\left( {Q {\bf E}_0^{(1)} \over \omega} \right)
{e^{- i \omega R_2/c } \over 2 i \sin \omega  \Delta \tau} .
\label{coeffads}
\ee
Substituting Eq.(\ref{coeffads}) into Eq.(\ref{modeexpads}), one obtains 
equation of motion for the string endpoint ${\bf X}_1(t)$ on the first 
D3-brane:
\be
M_{\rm w} 
\left( 1 + {R_1 \over R_2} \left( {\Delta \tau \partial_t
 \over
\tanh \Delta \tau \partial_t} - 1 \right) \right)^{-1}
\ddot {\bf X}_1 (t) \,\,\, = \,\,\, Q {\bf E}^{(1)} (t).
\label{ads1stend}
\ee
Comparison with the corresponding equation Eq.(\ref{leftend}) in the flat 
spacetime indicates that the infinite-order kernal has been changed 
completely both in functional form and in dependence on the D3-brane 
positions, $U_{1,2}$. Moreover, the kernel does not respect translational
invariance in $U$-coordinate as it depends on individual positions $U_{1,2}$ 
separately. If probed the anti-de Sitter spacetime using a static ruler such
as the W-boson, one would have concluded incorrectly the translational 
invariance from the W-boson mass, Eq.(\ref{adsmass}). Dynamic
information such as causal time-delay Eq.(\ref{adstimedelay}) does 
not exhibit translational invariance, either in `energy' or `scale' 
coordinates.
Despite these differences, the fact that the kernel makes the 
endpoint effectively heavier holds the same in anti-de Sitter spacetime
as well. 
 
Proceeding similarly, one also finds an equation of motion for the 
string endpoint ${\bf X}_2(t)$ on the second D3-brane: 
\be
M_{\rm w} \left( 
{\sinh \Delta \tau  \partial_t \over 
\Delta \tau \partial_t} \right)
\ddot {\bf X}_2(t) \,\,\,= \,\,\, Q {\bf E}^{(1)}(t).
\label{ads2ndend}
\ee
Interestingly, in this case, 
functional form of the infinite-order kernel is exactly the 
same as in the flat spacetime, Eq.(\ref{rightend}). However, argument
of the kernel, which will set the causal time-delay, is completely 
changed: in flat spacetime, it was $\Delta r = \ell_s^2 \Delta U$ proportional
to the W-boson mass, while, in anti-de Sitter spacetime, it is
$\Delta R = \Delta (g_{\rm eff} / U)$ and has no apparent functional relation
with the W-boson mass. As $R$ is the natural `size' variable in super 
Yang-Mills theory (see Eq.(\ref{uvir}), one again finds a hint from 
Eq.(\ref{ads2ndend}) that causality and locality of the anti-de Sitter
spacetime ought to be encoded in super Yang-Mills theory 
as a difference in size of brane waves.  
  
\subsubsection{Transmission Rate Across Open String}

It is straightforward to estimate the transmission rate of the brane-wave 
in anti-de Sitter spacetime. As argued above, the transmission has taken
place effectively via the stretched open string in the 
anti-de Sitter spacetime background. Compared to the flat spacetime,
the equation of motion Eq.(\ref{ads2ndend}) of the second endpoint depends
on $\Delta \tau = \Delta R/c$ (defined in Eq.(\ref{timedelay})), separation
between the two D3-branes measured in terms of `size' variable in the super 
Yang-Mills theory. Apart from this, as the equation of motion has exactly
the same form as in flat space, Eq.(\ref{rightend}), from Eq.(2.5), one finds
straightforwardly the transmission rate across F-string as
\be
T_{\rm F-string} = {2 \pi \over 3} \left( {Q^2 \over M_{\rm w} c^2}\right)^2
\left( 1 - 2 {\omega \over M_{\rm w} c^2} + \cdots \right)
\cdot
\left( 1 - \left( { \omega \over M_{\rm w} c^2} \right)^2 + \cdots \right)
\cdot
\left[ {\omega \Delta \tau \over \sin \omega  \Delta \tau } 
\right]^2.
\label{adsfstringtrans}
\ee
Again, the first two terms represent  classical and quantum results,
and the third summarizes radiation reactive force and virtual W-boson pair
effects. As the W-boson is a BPS state, its mass is protected in strong 
`t Hooft coupling limit. This explains for no change of the first three
terms from the result in flat spacetime. The last term, the string oscillation 
correction, is not protected by supersymmetry, even though its functional
form has remained the same, at least in the weak brane wave limit. 

Similarly, the transmission rate $T_{\rm D-string}$ for a D-string 
in anti-de Sitter spacetime is obtained as
\be
T_{\rm D-string} = {2 \pi \over 3} \left( {Q^2 \over  M_{\rm m} 
c^2}\right)^2
\left( 1 - 2 {\omega \over M_{\rm m} c^2} + \cdots \right)
\cdot
\left( 1 - \left( { \omega \over M_{\rm w} c^2} \right)^2 + \cdots \right)
\cdot
\left[ {\omega \Delta \tau \over \sin \omega
\Delta \tau } \right]^2.
\label{adsdstringtrans}
\ee
One again finds that the first three terms are protected against strong
`t Hooft coupling corrections, as both both magnetic monopole and W-boson
masses are BPS saturated.

It should be emphasized again that derivation of the above results is 
sufficiently controllable, 
despite the large-$N$ super Yang-Mills theory is in the
strong `t Hooft coupling regime. The expansion parameter $g_{\rm YM} 
{\bf E}_0/ M_{\rm w}$ is controllably small in so far as the brane wave is 
kept weak enough. 

\subsection{Locality and Causality in Anti-de Sitter Spacetime}

\subsubsection{Retarded Dynamics of String Endpoints}

As in flat spacetime, one can observe the causal time-delay, Eq.(\ref{timedelay}), in various ways. First, a pulse propagating along the sring in anti-de
Sitter spacetime is described by characteristic functions of the form
$g(t \pm R/c)$, see Eq.(\ref{modeexpads}). As such, it will take a causal
time delay $\Delta \tau$, as claimed in Eq.(\ref{timedelay}), for the pulse
to propagate across the two D3-branes. 
Alternatively, the equation of the second string endpoint, 
Eq.(\ref{ads2ndend}), written in an integral equation form:
\be
M_{\rm w} \ddot {\bf X}_2 (t) 
= \int dt' \Theta(t - t') 
\bigg\langle t \bigg\vert {\Delta \tau \partial_t \over 
\sinh \Delta \tau \partial_t} \bigg\vert t' \bigg\rangle
\cdot Q {\bf E}^{(1)}(t')
\label{retarded}
\ee
or in a difference equation form:
\be
{\bf X}_2 (t + \Delta \tau) - {\bf X}_2 (t - \Delta \tau)
= 2 \Delta \tau \int^t dt' \, Q {\bf E}^{(1)}(t')
\ee
displays manifestly the causal time-delay $\Delta \tau$ for brane-wave 
transmission. Agreement of the time-delay with the geometric distance,
Eq.(\ref{timedelay}) implies that the boundary interaction between 
F-string and D3-branes are local, even in the strong `t Hooft coupling limit.

Compared to that in flat spacetime, an important distinction of 
causal time-delay in anti-de Sitter spacetime is the fact that $\Delta \tau$
does {\sl not} depend on the sting scale at all. In flat spacetime, as 
$\Delta t = \ell_s^2 M_{\rm w}$, oscillation threshold and hence causal 
time-delay has disappeared completely in the supergraivty limit, 
$\ell_s \rightarrow 0$. This has now changed fundamentally in anti-de 
Sitter spacetime. The fact that oscillation threshold is finite, independent
of string scale, everywhere on the Coulomb branch moduli space suggests that
they will play an important role in defining the precise nature of the 
super Yang-Mills theory, to which we now turn. 

\subsubsection{Physical Thresholds in Anti-de Sitter Spacetime}
From the discussion of the last subsection, especially the transmission
rate formulae Eqs.(\ref{adsfstringtrans}, \ref{adsdstringtrans}), one finds
that two physical threshold scales in the super Yang-Mills theory are
`winding' threshold scale set by the W-boson mass and the `momentum' 
threshold scale set by the Yang-Mills `scale' size:
\be
\Lambda_{\rm w} = M_{\rm w}  = \Delta U\,; \qquad\qquad
\Lambda_{\rm m} =  
{1 \over \Delta \tau} = {U_1 U_2 \over g_{\rm eff}}  {1 \over \Delta U}.
\ee
As discussed already, the `winding' threshold scale $\Lambda_{\rm w}$ 
keeps the same value as in the flat spacetime, as it is protected by
supersymmetry. The `momentum' threshold scale $\Lambda_{\rm m}$, however,
has changed significantly. In particular, it has survived miraculously the
scaling limit, $\ell_s \rightarrow 0$, in obtaining the supergravity regime
in anti-de Sitter spacetime. In this sense, the stretched open string is a 
sort of non-critical string, whose tension is governed by a low-energy
scale in the super Yang-Mills theory.

A new uncertainty relation for anti-de Sitter spacetime is then given by
\be
\Lambda_{\rm w} \Lambda_{\rm m} \qquad \approx \qquad
{\cal O} \left( {U_1 U_2 \over g_{\rm eff}} \right).
\ee
Note that, in contrast to the flat spacetime situation, the uncertainty
relation does depend on the location of the two D3-branes, $U_{1,2}$, viz. 
location in the moduli space of the Coulomb branch. Thus, in any effects
related to causal retardation, the size of the threshold correction will
depend on the location, as $ \partial / \Lambda_{\rm m} \approx (g_{\rm eff} 
/ U_1 U_2) M_{\rm w} \partial$. For a fixed $M_{\rm w}$, the correction is
parametrically pronounced as the probe D3-branes zoom into the origin.

Stating differently, one finds that, as the `winding' threshold scale 
$\Lambda_{\rm w}$ is held fixed at a generic point in the 
Coulomb branch moduli space (away  from the origin), 
the `momentum' threshold $\Lambda_{\rm m}$ becomes smaller than 
$\Lambda_{\rm w}$ in the strong `t Hooft coupling limit, $g_{\rm eff}
\rightarrow \infty$. Causality and locality of the semi-classical supergravity
in anti-de Sitter spacetime will be then encoded to the super Yang-Mills
theory provided 
\be
\Lambda_{\rm w} \quad \gg \quad \Lambda_{\rm m}
\qquad \rightarrow \qquad
\Delta U \quad \gg \quad {\cal O} \left( \sqrt{U_1 U_2 \over g_{\rm eff}} \right).
\ee
As noted above, on top of the suppression in the strong `t Hooft coupling
limit, the condition is better satisfied near the origin of the Coulomb
branch moduli space. 

What about at the origin of the Coulomb branch? As the `momentum' threshold
depends both on $g_{\rm eff}$ and on $U_{1,2}$, the answer clearly depends on
the order one takes the limits $g_{\rm eff} \rightarrow \infty$ and
$U_{1,2} \rightarrow 0$. If one takes $g_{\rm eff} \rightarrow 0$ first 
and then $U_{1,2} \rightarrow 0$, it is clear that the `momentum' threshold
stays light all the time. If one takes oppositely, for example, $\Delta U
\rightarrow 0$ first and $U_{1,2} \rightarrow 0$ afterward, the `winding'
threshold always dominates and the non-critical string states will disappear
at the conformal point.   

\subsubsection{Holographic Encoding in Super Yang-Mills Theory}
We have now seen that, in large-$N$ super Yang-Mills theory, strong `t Hooft
coupling limit is what enables to encode causality and locality of the anti-de
Sitter spacetime. Let us now ask how the Thomson scattering process would
look like in the super Yang-Mills theory side. As the `momentum' threshold 
stays always lighter than the `winding' one, the W-boson is described more 
properly as an extended, bilocal object. Super Yang-Mills theory is not
the one defined either in conventional (dimension-four operators) or in
Dirac-Born-Infeld form, as they certainly do not contain such an object.

As the two endpoints can move 
independently, the two isospin components of the scattered brane waves can
carry the causal time-delay information. Let us see how this looks like
in the super Yang-Mills theory.   
The open string is now holographically projected to a shell of spherically 
symmetric Yang-Mills field configuration, as depicted in figure 10. The 
inner and outer radii are $R_1 = g_{\rm eff} / U_1$ and
$R_2 = g_{\rm eff} / U_2$, respectively. Once the W-boson is hit by the 
incident brane wave on the first D3-brane, concentric spherical waves of
the two isospin components will emanate. An observer sitting at spatial 
infinity on D3-branes will then perceive the radiation as originating from 
the same point in space (the center in figure 10), but with a time delay 
$\Delta \tau$. As is dictated by Eq.(\ref{timedelay}), the difference
in `size' coordinate of the wave-fronts in the super Yang-Mills theory
equals exactly to $\Delta \tau$. 

\begin{figure}[htb]
   \vspace{0cm}
   \centerline{
\epsfig{file=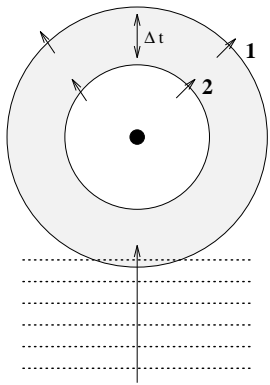,height=5cm,angle=0}
              }
\caption{Yang-Mills theory view of the Thompson scattering}
\end{figure}

\setcounter{equation}{0}

Having concluded that super Yang-Mills theory defined either in renormalizable
or in DBI Lagrangians is not capable of encoding causality and locality of
anti-de Sitter spacetime, we now ask what modifications, if possible, can 
do so. The answer to this question can be drawn again from Eq.(\ref{retarded}).
Expanding the infinite-order kernel for low-energy brane-wave, one finds
\bee
M_{\rm w} {\bf X}_2 (t) &=&
\sum_{n= -\infty}^{+\infty} (-)^n 
\int_{t \ge t'} dt' \, 
\bigg\langle t \bigg\vert {1 \over \partial_t^2 + (n / \Delta \tau)^2}
\bigg\vert t' \bigg\rangle \, Q{\bf E}^{(1)}(t') .
\nonumber \\
&=& 
Q \sum_{n = -\infty}^{+\infty} \sum_{m = 0}^\infty (-)^{m + n}
\left( {\Delta \tau \over n}\right)^{2m+2} {\cal O}_{2m} (t)
\eee
where
\be
{\cal O}_{2m} = \partial_t^{2m} {\bf E}(t).
\ee
Thus, the interaction of massive W-boson with the brane waves is governed by 
an infinite tower of higher-dimensional operators, ${\cal O}_{2m}$'s. Brane
waves can be gauge or Higgs fields, so there will be similar higher-derivative
operators including Higgs scalars. In $d=4, {\cal N}=4$ super Yang-Mills
theory, operators of these structures are classified as {\sl long 
supermultiplets.}

Perhaps, it should 
not be surpring that long supermultiplets are responsible for 
encoding causality and locality. After all, we have argued that string 
oscillator excitations are what is needed to for probing causality and
locality. It has been already known \cite{gubserklebanovpolyakov, witten}
that the long supermultiplets arise precisely from the massive string 
excitations, but the novelty of the above argument is that they can be
seen explicitly even at the level of single W-boson.

\newpage
\appendix
\leftline{\Large \bf Appendix: Effective Field Theory of String Endpoints}
\setcounter{equation}{0}
In this appendix, using covariant Green-Schwarz formalism, we will explain 
derivation of gauge-fixed, open string worldsheet action, 
Eqs.(\ref{flataction}, \ref{actionads}). 
The derivation turns out somewhat nontrivial, as two Majorana-Weyl spinors of 
the covariant Green-Schwarz action are constrained both by $\kappa$-symmetry
gauge-fixing and by open string boundary conditions. We will show that both
requirements lead to so-called `D-brane gauge' as a unique choice of the 
spinor projection. The `D-brane gauge' has been proposed previously as a
natural gauge-fixing condition of the $\kappa$ symmetry. Interestingly, we 
find that exactly the same condition also arises from the analysis of open 
string boundary conditions.  
 
We will also derive, after integrating out string excitations, an effective 
field theory describing low-energy dynamics of two endpoints of the string.
As pointed out in sections 2 and 4, the dynamics is described
by a Lagrangian with infinite-order kernels, which turns out exactly
the same as (two particle generalization of) the Pais-Uhlenbeck's model. 
Remarkably, according to the analysis of Pais and Uhlenbeck, the particular
infinite-order kernel is the one compatible with convergence, 
positive-definiteness and causality.  

\section{Green-Schwarz Action in Flat Spacetime}
Dynamics of an open F- or D-string stretched between the two parallel 
D3-branes, as depicted in Figure 2, is governed by the covariant Green-Schwarz 
action \cite{greenschwarz}. Denote bosonic and fermionic string coordinates
as $(X^M; \Theta^A_\alpha)$ 
$(M = 0, \cdots, 9; A = 1, 2; \alpha = 1, \cdots, 32)$, which map 
worldsheet $\Sigma$ (spanned by $\sigma^i$ $(i=0,1)$) to coset superspace 
$Poincare(9,1 \vert 2) / SO(9,1)$. 
In Type IIB string theory, $\Theta^{1,2}$ 
are Majorana-Weyl spinors of same chirality. 
In the Nambu-Goto formulation, the worldsheet action is given by 
\bee
I_{\rm string} &=& I_{\rm GS} + I_{\rm D3-Q}
\nonumber \\
I_{\rm GS}
&=& - T \int_\Sigma d\tau d \sigma \,
\Big[ \sqrt{ - {\rm det} L_i^{\hat a} L_j^{\hat a}}
+ i \int_{M_3} L^{\hat a} \wedge \overline{L}^I \tau_3^{IJ} 
\Gamma^{\hat a} \wedge L^J.
\label{flatgsaction}
\eee
Here, $I_{\rm D3-Q}$ is a boundary action of the string endpoints, $M$ 
denotes a three-dimensional manifold whose boundary equals to
the string worldsheet, and  
\bee
L^{\hat a} &=& \delta_M^{\hat a} d X^M - i \overline{\Theta}^A 
\Gamma^{\hat a} d \Theta^A , \qquad (\hat a = 0, 1, \cdots, 9)
\nonumber \\
L^I &=& d \Theta^I.
\eee
are invariant Cartan one-forms of the coset superspace.
The Wess-Zumino term in Eq.(\ref{flatgsaction}) is given by
\be
L_{\rm WZ} = 
2i d X^{\hat a} \!\wedge \!\left( \overline{\Theta}^1 \Gamma^{\hat a} 
d \Theta^1 -  \overline{\Theta}^2 \Gamma^{\hat a} d \Theta^2  \right) 
-  2 \left(\overline{\Theta}^1 \Gamma^{\hat a} d \Theta^1\right) 
\!\wedge\! 
\left(\overline{\Theta}^2 \Gamma^{\hat a} d \Theta^2 \right) .
\ee

Apart from the boundary Lagrangian, the covariant Green-Schwarz action 
Eq.(\ref{flatgsaction}) is invariant under ${\cal N}=2$ global 
supersymmetry ($A = 1, 2$):
\be
\delta_\epsilon \Theta^A = \epsilon^A;  
\qquad 
\delta_\epsilon X^{\hat a} = i \overline \epsilon^A \Gamma^{\hat a} \Theta^A,
\ee
and local $\kappa$ symmetry, when expressed in terms of worldsheet {\sl 
scalar} spinors $\tilde \kappa$ \footnote{The worldsheet
vector $\kappa$ and worldsheet scalar $\tilde \kappa$ are related each other
by $\kappa^{1,2}_i = - i \Gamma^{\hat a} \Pi^{\hat a}_i 
\tilde \kappa^{2,1}$.}, 
\be
\delta_{\tilde \kappa} \, \Theta^A = \left( 1 + (-)^A {\cal P} \right)
\tilde \kappa^A,  \qquad
\delta_\kappa \, X^{\hat a} = i \overline{\Theta}^A \Gamma^{\hat a} 
\delta_{\tilde \kappa} \Theta^A,
\ee 
where
\be
{\cal P} = {1 \over 2} {L^{\hat a} \wedge L^{\hat b} \over \sqrt{-h}} 
\Gamma^{\hat a \hat b}, \qquad  {\cal P}^2 = 1.
\ee

\subsection{Open String Boundary Conditions}
The last term in Eq.(\ref{flatgsaction}) is the Lagrangian describing 
interaction of the open string endpoints with gauge and Higgs fields
on the worldvolume of the two D3-branes. Denoting the $d=4, {\cal N}=4$
vector supermultiplet as $(A, \lambda)$, the boundary Lagrangian in the
lowest order in derivative expansion reads
\be
L_{\rm D3-Q} = Q \oint_{\partial \Sigma}
d \tau \dot X^{\hat a} \left( A_{\hat a}(X) - \overline \lambda (X)
\Gamma_{\hat a} \theta \right).
\ee
The boundary action represents turning on gauge and Higgs fields on the
D3-brane worldvolume. 
The boundary conditions, derived from variation of the Green-Schwarz action,
are given by \footnote{Once, on the D3-brane worldvolume, Higgs and gauge
fields are turned on, boundary conditions are expected to be modified. It is
easy to show, however, that Higgs field background does not modify the 
Dirichlet
boundary condition at all, while gauge field background affects the Neumann
boundary condition. As we will be considering background fields that are 
weak enough, these modifications will be ignored throughout.}
\bee
D\quad &:& \qquad 
\delta X^\perp = 0, \qquad 
\overline \Theta^1 \Gamma^\perp \delta \Theta^1 +
\overline \Theta^2 \Gamma^\perp \delta \Theta^2 = 0
\label{d} \\
N \quad &:& \qquad
\Pi^{\parallel}_\sigma = 0, \qquad
\overline \Theta^1 \Gamma^{\parallel} \delta \Theta^1 
- \overline \Theta^2 \Gamma^{\parallel} \delta \Theta^2 = 0,
\label{n}
\eee
for Dirichlet and Neumann directions, respectively.
To proceed further with gauge fixing, one will need to find linear 
boundary conditions for fermions that are consistent with Eqs.(\ref{d},
\ref{n}).
Because of sign difference between $A=1,2$ terms in Eqs.(\ref{d}, \ref{n}),
a unique choice of the linear boundary condition is of the form
\be
\Theta^A = \Gamma_5 {{\cal M}^A}_B \Theta^B, \qquad
\Gamma_5 \equiv \Gamma^0 \Gamma^1 \cdots \Gamma^3,
\label{linear}
\ee
where $(2 \times 2)$ matrix ${\cal M}$ is determined to be ${\cal M}_{AB}
= \epsilon_{AB}$ by 
the requirement
that the boundary condition is compatible with Majorana and Weyl conditions
and with invertibility of Eq.(\ref{linear}). 
Hence, for open Green-Schwarz superstring attached between the two D3-branes,
the boundary conditions are:
\bee
\delta X^\perp = 0, &\qquad& \partial_\sigma X^{\parallel} = 0
\nonumber \\
\left(\delta_{AB} - \Gamma_5 \epsilon_{AB} \right) \Theta^B = 0
&\quad&
\left(\delta_{AB} + \Gamma_5 \epsilon_{AB} \right) \partial_\sigma
\Theta^B = 0.
\eee

\subsection{Gauge Fixing}
Due to the local $\kappa$-symmetry, half of the degrees of freedom of 
$\Theta^{1,2}$'s are redundant.  Since the spinors $\Theta^{1,2}$ have the 
same chirality, the $\kappa$-symmetry may be gauge fixed conveniently by 
setting $\Theta^1 = \Gamma_5 \Theta^2 \equiv {1 \over 2} \Psi$.
Using equation of motion for $L^{\hat a}$'s, a straightforward calculation 
yields the $\kappa$-gauge-fixed action
\be
I_{\rm GS} = T  \int\!\!\!\int d\tau d \sigma \,
\left[ \sqrt{ - {\rm det} \partial_i X^{\hat a} \partial_j X^{\hat a} }
+ { i \over 2} d X^{\hat a} \wedge 
(\overline{\Psi} \Gamma^{\hat a} d \Psi) \, \right]. 
\ee
In the gauge-fixed action, 
half of ${\cal N}=2$ supersymmetry is realized linearly
\be
\delta_\epsilon X^{\hat a} = i \overline{\epsilon} \Gamma^{\hat a} \Psi, 
\qquad
\delta_\epsilon \Psi = - {1 \over 2}    
{d X^{\hat a} \wedge d X^{\hat b} \over \vert\!\!\vert d X \wedge d X
\vert \!\! \vert} \Gamma^{\hat a \hat b} \epsilon,
\ee
and the other half as a nonlinear fermionic symmetry
\be
\delta_\xi X^{\hat a} = 0, \qquad \delta_\xi \Psi = \xi.
\ee

For the reparametrization invariance, we will choose the static gauge
$\sigma^0 = t$, $\sigma^1 = r$. After expanding the action to quadratic 
order and reinstating the speed of light $c$, one obtains the 
gauge-fixed action of the open string as:
\be
I_{\rm GS} = I_{\rm string} + I_{\rm D3-Q}
\ee
where
\bee
I_{\rm string} &=& \int \!dt \, \left[ - M_{\rm w} 
+ {T \over 2} \int_{r_1}^{r_2} dr \, 
\left( {1 \over c^2} (\partial_t {\bf X})^2
- (\partial_r {\bf X})^2 \right) + \cdots \right]
\label{flataction2} \\
\nonumber \\
I_{\rm D3-Q} &=& \!\!\!\! \sum_{I = 1,2}
 \int dt \, Q_{\rm I} \left[ A_0^{\rm (I)}({\bf r}, t) 
+ {1 \over c} \dot {\bf X}_{\rm I}(t) \cdot {\bf A}^{\rm (I)} ({\bf r}, t)
\right]_{{\bf r} = {\bf X}_{\rm I}(t)}, \qquad (Q_1 = - Q_2 = \pm 1).
\nonumber \\
\eee

\subsection{Low-Energy Effective Lagrangian of String Endpoints}
We will begin with solving the equations of motion for ${\bf X}(t,r)$, subject
to fixed but arbitrary location of the string endpoints, ${\bf X}_{1,2} (t)$. 
Consider a Fourier transformed, harmonic solution 
\bee
\tilde{\bf X} (\omega, r)
&=& \int dt e^{ + i \omega t} {\bf X}(t, r)
\nonumber \\ 
&=&  {\bf a}(\omega) e^{-i \omega (t - r/c)} 
+ \tilde{\bf a}(\omega) e^{- i \omega(t + r/c)}. 
\nonumber
\eee
The boundary condition specified,
\be
\tilde {\bf X}(\omega, r) \bigg\vert_{r = r_1} = 
\tilde{\bf X}_1 (\omega), \qquad
\tilde{\bf X}(\omega, r) \bigg\vert_{r=r_2} = 
\tilde{\bf X}_2 (\omega),
\nonumber
\ee
relates 
the spectral amplitudes ${\bf a}(\omega), \tilde{\bf a}(\omega)$ uniquely 
to the (Fourier transformed) string endpoints $\tilde{\bf X}_{1,2}(\omega)$.
Using the relation, it is possible to express normal derivatives at the 
string endpoints
\be
\partial_r \tilde {\bf X}_1(\omega) 
\equiv \partial_r \tilde{\bf X}(\omega, r) \bigg\vert_{r = r_1},
\qquad
\tilde {\bf X}_2(\omega) \equiv \partial_r \tilde{\bf X}
(\omega, r) \bigg\vert_{r = r_2}
\nonumber
\ee
in terms of $\tilde {\bf X}_{1,2}(\omega)$. 
The result is ($\Delta t \equiv \Delta r / c$)
\bee
\partial_r \tilde{\bf X}_1 (\omega) 
= - {\omega \Delta t \over \sin \omega \Delta t}
\left( { \tilde {\bf X}_1 (\omega) \cos \omega \Delta t - \tilde {\bf X}_2 
(\omega) \over c \Delta t} \right)
\nonumber \\
\partial_r \tilde{\bf X}_2 (\omega)
= - {\omega \Delta t \over \sin \omega \Delta t}
\left( { \tilde{\bf X}_1 (\omega) - \tilde{\bf X}_2 (\omega) 
\cos \omega \Delta t \over c \Delta t} \right).
\label{xderivative}
\eee 

In order to obtain an effective action for the string endpoints, we will
need to integrate out massive string excitations. As the string action,
$I_{\rm string}$, is quadratic in ${\bf X}(r, t)$ at leading order, it 
amounts to imposing the equation of motion of ${\bf X}(r, t)$ back to the
action, Eq.(\ref{flataction2}). One then obtains, after Fourier-transform
back to ${\bf X}_{1,2}(t)$,
\be
I_{\rm string} = \int \! dt \, \, \left[ - M_{\rm w}
- {T \over 2} \Big({\bf X}_2 (t) \cdot \partial_r {\bf X}_2 (t) - 
{\bf X}_1 (t) \cdot \partial_r {\bf X}_1 (t) \Big) + \cdots \right] .
\label{reduced1}
\ee
The string action is now expressed entirely in terms of boundary data of
the open string positions on the two D3-branes. 

Inserting Eq.(\ref{xderivative}) to Eq.(\ref{reduced1}), one finally obtains 
the low-energy effective action of the string endpoints:
\bee
I_{\rm GS} &=& I_{\rm string} + I_{\rm D3-Q}
\label{flatbdryaction} \\
I_{\rm string} &=& \int dt \, \left[ -M_{\rm w} -{T \over 2} \sum_{I,J = 1,2}
{\bf X}_I (t) {\cal K}^{\rm flat}_{IJ} 
\left( \Delta t \partial_t \right) {\bf X}_J (t) \,\, \right]
\nonumber \\
I_{\rm D3-Q}  
&=& \sum_{\rm I = 1, 2} \int dt \, Q_{\rm I} \left[ A_0^{\rm (I)} 
({\bf r}, t) + 
{1 \over c} \dot {\bf X}_{\rm I}(t) \cdot {\bf A}^{\rm (I)} 
({\bf r}, t) \right]_{{\bf r} = {\bf X}_{\rm I} (t)},
\qquad (Q_1 = - Q_2 = \pm 1).
\nonumber
\eee
In the action, $I_{\rm string}$, the infinite-order kernels are defined by:
\bee
{\cal K}^{\rm flat}_{11} = {\cal K}^{\rm flat}_{22} &=& + {1 \over \Delta r}
\left( {\Delta t \partial_t \over \tanh \Delta t \partial_t} \right)
\nonumber \\
{\cal K}^{\rm flat}_{12} = {\cal K}^{\rm flat}_{21} &=& 
- {1 \over \Delta r} 
\left( {\Delta t \partial_t \over \sinh \Delta t \partial_t} \right).
\eee

Let us now focus on the Thomson scattering by D3-brane waves, studied in section 2. Initially, the radiation field is present only for the first D3-brane, 
viz. $A_0^{(2)} = {\bf A}^{(2)} = 0$. 
In this case, the equation of motion for ${\bf X}_2(t)$ is reduced 
simply to
\be
 \left( \cosh \Delta t \partial_t\right) \,  {\bf X}_2 (t) = {\bf X}_1 (t),
\label{x2eqn}
\ee
viz. a retardation relation
\be
{\bf X}_2 (t) = \int dt' \bigg\langle t \bigg\vert 
{1 \over \cosh \Delta t \partial_t} \bigg\vert t' \bigg\rangle \, 
{\bf X}_1 (t').
\ee
After ${\bf X}_2(t)$ is integrated out of $I_{\rm string}$ 
using Eq.(\ref{x2eqn}), the low-energy effective Lagrangian for 
the Thomson scattering process is obtained:
\bee
L_{\rm Thomson}
= &-& M_{\rm w} + {M_{\rm w} \over 2} \dot {\bf X}_1(t) \left(
{\tanh \Delta t \partial_t \over \Delta t \partial_t} \right) 
\dot{\bf X}_1 (t) + \cdots \nonumber \\
\nonumber \\
&+& Q \left[ A_0^{(1)} ({\bf r}, t) 
+ {1 \over c} \dot {\bf X}_1 \cdot {\bf A}^{(1)} ({\bf r}, t) \right]_{ 
{\bf r} = {\bf X}_1 (t)}.
\label{thomsonlag}
\eee
The Eqs.(\ref{leftend}, \ref{rightend}) follow immediately from the above
effective Lagrangian and the retardation relation Eq.(\ref{x2eqn}).

\section{Green-Schwarz Action in Anti-de Sitter Spacetime}
The Green-Schwarz action in $AdS_5 \times S^5$ can be obtained in a closed
form via super-coset space construction \cite{tseytlin} \footnote{ 
For a D-string, the action can be constructed similarly \cite{parkrey}. }.
In Nambu-Goto form, the action reads
\be
I_{\rm AdS}
= T \int\!\!\!\int d\tau d \sigma \,
\left[ \sqrt{-{\rm det} L_i^{\hat a} L_j^{\hat a} } + 
4i \int_0^1 ds L_i^{\hat a} (s) \overline{\Theta} \tau_3 
\Gamma^{\hat a} L_j (s) \, \right].
\ee
Here, ${\hat a} = (a, a') = (0,1, \cdots, 4, 5, \cdots, 9)$, $I,J = 1,2$
and $\tau_{1,2,3}$ are Pauli matrices. The invariant worldsheet one-forms
$L^I(s)$ and $L^{\hat a}(s)$ are given by
\bee
L^I(s) &=& \left( {\sinh s {\cal M} \over {\cal M}} {\cal D} \Theta \right)^I
\nonumber \\
L^{\hat a}(x) &=& e^{\hat a}_M d X^M - 4i \overline{\Theta}^I \Gamma^{\hat a}
\left( {\sinh^2 {1 \over 2} s {\cal M} \over {\cal M}^2} 
{\cal D} \Theta \right)^I,
\eee
where $(X^M, \Theta^I)$ are  the bosonic and fermionic F-string coordinates and
\bee
\left({\cal M}^2 \right)^{IJ} &=&
\left[ \gamma^{a'} (i \tau_2 \Theta)^I \overline{\Theta}^J \gamma^{a'}
- \gamma^{a} (i \tau_2 \Theta)^I \overline{\Theta}^J \gamma^{a} \right]
+ {1 \over 2}
\left[ \gamma^{a'b'}  \Theta^I ( \overline{\Theta} i \tau_2)^J \gamma^{a'b'} 
- 
\gamma^{ab} \Theta^I (\overline{\Theta} i \tau_2)^J \gamma^{ab} \right]
\nonumber \\
({\cal D} \Theta)^I
&=&
\left( \partial + {1 \over 4} \omega^{\hat a \hat b} \Gamma_{\hat a \hat b}
 \right) \Theta^I + {1 \over 2} E^{\hat a} \Gamma_{\hat a}
\left( \tau_2 \Theta \right)^I .
\eee
Also, $L^I \equiv L^I (s =1)$, $L^{\hat a} \equiv L^{\hat a}(s=1)$.

The action is invariant under the ${\cal N}=2$ global supersymmtry 
and local $\kappa$-symmetry. Expressing again in terms of worldsheet
scalar spinors, the symmetry transformations rules are essentially
the same as Eqs.(A.5, 6), except now the indices are repeated over
$AdS_5$ and $S^5$.  
:

\subsection{Gauge Fixing}

In order to fix the local $\kappa$-symmetry, it turns out most convenient to 
make a change of variable $R = g_{\rm eff} / U$ and use the Poincar\'e 
coordinates.
Up to the conformal rescaling, the ten-dimensional spacetime is flat. 
To proceed in an analogous manner to the flat spacetime situation,  
we will follow the observation of Kallosh \cite{kallosh1, kallosh2, 
kallosh3} and take the gauge-fixing
\footnote{Different gauge-fixing choice has been advocated recently
\cite{rajaramanrozali}.}
\be
\Theta^I_- = 0 \qquad {\rm where}
\qquad 
\Theta^I_\pm \equiv ({\cal P}_\pm \Theta)^I
, \quad {\cal P}_\pm^{IJ} 
= {1 \over 2} \left( \tau_0 \pm i \Gamma_5 \tau_2 \right)^{IJ}.
\ee

The $\kappa$-gauge-fixed action is quartic in fermions. 
However, if one makes a change of variable to the `size' variable,
 $R = g_{\rm eff}/U$, as mentioned in seciton 4, the quartic fermon terms 
are turned into quadratic ones only. Moreover, in the $R$-coordinates,
as we will see below, the reparametrization gauge-fixed action 
also simplifies considerably \footnote{In 
\cite{kallosh3}, closely related observation has been made
but was interpreted as the effect of T-duality along all directions of
D3-brane worldvolume. According to the present argument, this 
interpretation seems unnecessary.}.

Again, choosing the static gauge $\sigma^0 = t$, $\sigma^1 = R$, 
one finds that the gauge-fixed Green-Schwarz action of the open string
in anti-de Sitter spacetime is given by:
\be
I_{\rm AdS} =  \left( L_{\rm string} + L_{\rm D3-Q} \right)
\ee
where
\bee
I_{\rm string} 
&=& \int dt \, \left[ - M_{\rm w} 
+ {g_{\rm eff} \over 2}  \int_{R_1}^{R_2} {d R \over R^2}
\left( {1 \over c^2} (\partial_t {\bf X})^2 - (\partial_R {\bf X})^2 \right)
+ \cdots \right]
\label{adsbulk} \\
\nonumber \\
L_{\rm D3-Q} &=& \!\!\! \sum_{I = 1, 2} \! \int dt \, 
Q_{\rm I} \left[ A_0^{\rm (I)} ({\bf r}, t)
+ {1 \over c} \dot {\bf X}_{\rm I} (t) \cdot {\bf A}^{\rm (I)}
({\bf r}, t) \right]_{{\bf r} = {\bf X}_{\rm I} (t)},
\quad (Q_1 = - Q_2 = \pm 1).
\label{adslagrangian}
\eee
and the boundary interaction action is exactly the same in the flat 
spacetime.  
\subsection{Low-Energy Effective Lagrangian of String Endpoints}
One can obtain low-energy effective Lagrangian of the string endpoints
repeating steps of flat spacetime case, section A.2. 
After Fourier transformation, a harmonic solution of the string coordinate 
$\tilde {\bf X}(\omega, R)$ now takes 
a form
\bee
\tilde{\bf X}(\omega, R) &=& \int dt e^{+ i \omega t} {\bf X}(t, R)
\nonumber \\
&=& {\bf a}(\omega) \left( 1 - i \omega {R \over c} \right)
e^{-i \omega (t - R/c)} + \tilde{\bf a}(\omega) 
\left( 1 + i \omega {R \over c} \right)
e^{-i \omega (t + R/c)},
\nonumber
\eee
where the harmonic modes are expanded in terms of the Hankel functions, 
$H^{(\pm)}_{3/2}(\omega R)$.
Eliminating ${\bf a}(\omega), \tilde{\bf a}(\omega)$ between 
$\tilde{\bf X}_{1,2}(\omega)$ and $\partial_z \tilde{\bf X}_{1,2}(\omega)$, 
one finds relations $(\Delta \tau \equiv \Delta z / c)$
\bee
{1 \over R_1} \partial_R \tilde {\bf X}_1 (\omega) 
&=&
{\omega^2 \over {\cal N}}
\left[ \left( \sin \omega \Delta \tau + \omega {R_2 \over c} \cos \omega
\Delta \tau \right) \tilde{\bf X}_1 (\omega) 
- \omega {R_1 \over c} \tilde {\bf X}_2 (\omega) \right]
\nonumber \\
{1 \over R_2} \partial_R \tilde {\bf X}_2 (\omega)
&=&
{\omega^2 \over {\cal N}}
\left[ \omega {R_2 \over c} \tilde {\bf X}_1 (\omega) + 
\left( \sin \omega \Delta \tau - 
\omega {R_1 \over c} \cos \omega \Delta \tau \right) \tilde {\bf X}_2 (\omega)
 \right],
\label{relations}
\eee
where
\be
{\cal N} = \left( 1 + \omega^2 {R_1 R_2 \over c^2} \right) \sin \omega 
\Delta \tau - \omega \Delta \tau \cos \omega \Delta \tau .
\nonumber
\ee
Imposing the equation of motion for ${\bf X}(t, z)$ to the action
Eq.(\ref{adsbulk}), one obtains
\be
I_{\rm string}
= \int dt \, \left[
- M_{\rm w} - {g_{\rm eff} \over 2} 
\left( {1 \over R_2^2} {\bf X}_2 (t) \cdot \partial_R {\bf X}_2 (t)
- {1 \over R_1^2} {\bf X}_1 (t) \cdot \partial_R {\bf X}_1 (t) \right)
+ \cdots \right].
\label{reduced}
\ee
Inserting Eq.(\ref{relations}) to the action Eq.(\ref{reduced}), after
Fourier-transforming back, one finally arrives at the low-energy effective 
action of the string endpoints in anti-de Sitter spacetime:
\bee
I_{\rm AdS} &=&  \left( I_{\rm string} + I_{\rm D3-Q} \right)
\label{furtherreduced} \\
I_{\rm string} &=& \int dt \, \left[ -M_{\rm w} - {g_{\rm eff} \over 2} 
\sum_{I, J = 1, 2} {\bf X}_I(t) \cdot {\cal K}_{IJ}^{\rm AdS} 
(\Delta \tau \partial_t) \cdot {\bf X}_J (t) + \cdots \right]
\nonumber \\
\nonumber \\
I_{\rm D3-Q} &=& \sum_{I=1,2} \int dt \,
Q_{\rm I} \left[ A_0^{\rm (I)} ({\bf r}, t) + {1 \over c}
\dot {\bf X}_{\rm I}(t) \cdot {\bf A}^{\rm (I)} ({\bf r}, t) 
\right]_{{\bf r} = {\bf X}_{\rm I}(t)},
\qquad (Q_1 = - Q_2 = \pm 1).
\nonumber\\
\nonumber 
\eee
The infinite-order kernels in the action $I_{\rm string}$ are now given by:
\bee
{\cal K}_{11}^{\rm AdS} &=& - {1 \over R_1} 
\left( 1 + {R_2/c \over \Delta \tau}  
{ \Delta \tau \partial_t \over \tanh \Delta \tau \partial_t} \right) 
\left( 1 - {R_1 R_2 \over c^2} \partial_t^2 - { \Delta \tau \partial_t 
\over \tanh \Delta \tau \partial_t} \right)^{-1}
\nonumber \\
\nonumber\\
{\cal K}_{22}^{\rm AdS} &=& + {1 \over R_2} 
\left( 1  - {R_1/c \over \Delta \tau} 
{ \Delta \tau \partial_t \over \tanh \Delta \tau \partial_t} \right)
\left(1 - {R_1 R_2 \over c^2} \partial_t^2 - {\Delta \tau \partial_t 
\over \tanh \Delta \tau \partial_t} \right)^{-1}
\nonumber \\
\nonumber \\
{\cal K}_{12}^{\rm AdS} &=& {\cal K}_{21}^{\rm AdS} = 
+ {1 \over \Delta R}
\left( {\Delta \tau \partial_t \over \sinh \Delta \tau \partial_t} \right)
\left( 1 - {R_1 R_2 \over c^2} \partial_t^2 - {\Delta \tau \partial_t \over
\tanh \Delta \tau \partial_t} \right)^{-1}.
\label{adskernels}\\
\nonumber
\eee

Let us again consider Thomson scattering of D3-brane wave, studied in section
4, by setting $A_0^{(2)} = {\bf A}^{(2)} = 0$. From Eq.(\ref{furtherreduced}),
equation of motion for ${\bf X}_2(t)$ yields a retardation relation 
\be
\left( 1 +  {R_1 \over R_2} 
\left( {\Delta \tau \partial_t \over \tanh \Delta \tau \partial_t} - 1 \right) 
\right)
\left( {\sinh \Delta \tau \partial_t \over \Delta \tau \partial_t} \right)
\, {\bf X}_2 (t) = {\bf X}_1 (t).
\label{adsrel}
\ee
Integrating out ${\bf X}_2(t)$ from the action $I_{\rm string}$, using 
Eq.(\ref{adsrel}), one finally obtains the low-energy effective Lagrangian
for the Thomson scattering of D3-brane waves:
\bee
L_{\rm Thomson} = &-& M_{\rm w} 
+ {M_{\rm w} \over 2} \dot {\bf X}_1(t) 
\left( 1 + {R_1 \over R_2} \left( {\Delta \tau \partial_t \over 
\tanh \Delta \tau \partial_t} - 1 \right) \right)^{-1} \dot{\bf X}_1 (t)
+ \cdots
\nonumber \\
&+& Q \left[ A^{(1)}_0({\bf r}, t)
+ {1 \over c} 
\dot {\bf X}_1(t) \cdot {\bf A}^{(1)}({\bf r}, t) \right]_{{\bf r}
= {\bf X}_1(t)}.
\label{adsefflag}
\eee
From the effective Lagrangian Eq.(\ref{adsefflag}) and the retardation relation
Eq.(\ref{adsrel}), one gets easily the equations of motion of string
endpoints, Eqs.(4.44, 4.45). 
 

\vfill
\eject

\end{document}